\documentclass[11pt,a4paper]{article}
\pdfoutput=1

\usepackage{jheppub}

\usepackage{amssymb}
\usepackage{amsfonts}
\usepackage{graphicx}
\usepackage{epstopdf}
\usepackage{dcolumn}
\usepackage{amsmath}
\usepackage{latexsym,bm}
\usepackage{amsthm}
\usepackage{slashed}
\usepackage{float}
\usepackage{color}
\usepackage{url}
\usepackage{longtable}
\usepackage{subfig}

\newcommand \xoverline[2][0.75]{
    \sbox{\myboxA}{$\m@th#2$}
    \setbox\myboxB\null
    \ht\myboxB=\ht\myboxA
    \dp\myboxB=\dp\myboxA
    \wd\myboxB=#1\wd\myboxA
    \sbox\myboxB{$\m@th\overline{\copy\myboxB}$}
    \setlength\mylenA{\the\wd\myboxA}
    \addtolength\mylenA{-\the\wd\myboxB}
    \ifdim\wd\myboxB<\wd\myboxA
       \rlap{\hskip 0.5\mylenA\usebox\myboxB}{\usebox\myboxA}%
    \else
        \hskip -0.5\mylenA\rlap{\usebox\myboxA}{\hskip 0.5\mylenA\usebox\myboxB}%
    \fi}
\makeatother

\newcommand{\ba}{\begin{aligned}}
\newcommand{\ea}{\end{aligned}}
\def \be {\begin{equation}}
\def \ee {\end{equation}}
\def \bsp {\begin{split}}
\def \esp {\end{split}}
\def \bea {\begin{eqnarray}}
\def \eea {\end{eqnarray}}

\def\mc{\mathcal}

\def\mb{\mathbb}

\def \bp{\begin{pmatrix}}
\def\ep{\end{pmatrix}}

\title{On the Elliptic Calabi-Yau Fourfold with Maximal $h^{1,1}$}

\author[a]{Yi-Nan Wang}

\affiliation[a]{Mathematical Institute, University of Oxford, \\
Andrew-Wiles Building,  Woodstock Road, Oxford, OX2 6GG, UK}

\emailAdd{yinan.wang@maths.ox.ac.uk}

\preprint{\today \hspace*{0.1in} }

\abstract{In this paper, we explicitly construct the smooth compact base threefold for the elliptic Calabi-Yau fourfold with the largest known $h^{1,1}=303\,148$. It is generated by blowing up a smooth toric ``seed'' base threefold with $(E_8,E_8,E_8)$ collisions. The 4d F-theory compactification model over it has the largest geometric gauge group, $E_8^{2\,561}\times F_4^{7\,576}\times G_2^{20\,168}\times SU(2)^{30\,200}$, and the largest number of axions, 181\,820, in the known 4d $\mc{N}=1$ supergravity landscape. We also prove that there are at least $1100^{15\,048}\approx 7.5\times 10^{45\,766}$ different flip and flop phases of this  base threefold. Moreover, we find that many other base threefolds with large $h^{1,1}$ in the 4d F-theory landscape can be constructed in a similar way as well.}


\begin{document}

\maketitle

\section{Introduction}

In recent years, there have been a lot of activities to determine the boundary of the string landscape, see~\cite{Brennan:2017rbf} for a brief overview. In even space-time dimensions, F-theory compactification models~\cite{Vafa:1996xn,Morrison:1996na,Morrison:1996pp,Weigand:2010wm,Weigand:2018rez} give rise to a large class of supersymmetric vacuum solutions. In fact, the F-theory geometric landscape has the largest known number of vacuum solutions, along with the largest gauge groups and the largest number of moduli fields on a particular geometry.

For 6d $(1,0)$ supergravity from F-theory on an elliptic Calabi-Yau threefold $X_3$, the set of base complex surfaces has been studied in \cite{Morrison:2012np,Morrison:2012js,Taylor:2012dr,Martini:2014iza,Taylor:2015isa}. Especially, the total number of 2d toric base surfaces is computed to be 61\,539~\cite{Morrison:2012js}, under the condition that the generic elliptic fibration over them does not have any non-flat fiber over toric points\footnote{In the Weierstrass model $y^2=x^3+fx+g$, the condition is that $(f,g)$ does not vanish to order $(4,6)$ or higher at toric points.}. The different fibrations over each base have been explored in e. g. \cite{Morrison:2012ei,Johnson:2014xpa,Klevers:2016jsz,Johnson:2016qar,Klevers:2017aku,Huang:2018gpl,Raghuram:2018hjn,Taylor:2019ots}. However, the total number of elliptic Calabi-Yau threefolds has not been estimated yet.

In particular, the elliptic Calabi-Yau threefold with $(h^{1,1},h^{2,1})=(491,11)$ has the largest known number of K\"{a}hler moduli, along with the largest geometric gauge group:
\be
G=E_8^{17}\times F_4^{16}\times G_2^{32}\times SU(2)^{32}
\ee
in the known 6d (1,0) supergravity landscape. It also has the largest $h^{1,1}$ in the known set of compact Calabi-Yau threefolds~\cite{Kreuzer:2000xy}.

For 4d $\mc{N}=1$ supergravity from F-theory on an elliptic Calabi-Yau fourfold $X_4$, even the set of toric base threefolds has not been fully classified. If we allow bases that only support non-flat fibration over complex curves, then the lower bound of such toric threefolds was proved to be $\frac{4}{3}\times 2.96\times 10^{755}$ in \cite{Halverson:2017ffz}, and further estimated to be $\sim 10^{3,000}$ in \cite{Taylor:2017yqr}. In \cite{Taylor:2017yqr}, the notion ``good base'' was introduced to describe the subset of bases that  support a flat fibration over complex curves. This fibration cannot have a complex three-dimensional fiber over a point either. In the language of Weierstrass polynomials, it is required that $(f,g)$ do not vanish to order $(4,6)$ or higher on any complex curve, and $(f,g)$ do not vanish to order $(8,12)$ or higher over any point. Such bases are convenient for the computation of $h^{1,1}(X_4)$, because the Tate-Shioda-Wazir formula
\be
h^{1,1}(X_4)=h^{1,1}(B_3)+\mathrm{rank}(G)+1
\ee
can be applied. Physically, a good base can support a 4d supergravity description, without strongly coupled matter sectors localized on the non-minimal loci\footnote{In the definition of a good base, we allow points where $(f,g)$ vanish to order $(4,6)$ or higher. These points can give rise to higher order Yukawa coupling terms~\cite{Achmed-Zade:2018idx}.}. The lower bound of the total number of good bases is estimated to be $\sim 10^{48}$ for the bases with $h^{1,1}<130$~\cite{Taylor:2015ppa} and $\sim 10^{250}$ for the bases with $h^{1,1}>1000$~\cite{Taylor:2017yqr}. It was also found that the $h^{1,1}$ of such good bases are  concentrated at certain discrete values. Hence it was postulated that non-trivial structures exist on this subset.

These large exponential numbers arise from the large number of flip and flop operations on the base threefold, which potentially leads to different matter curves and 4d low energy physics~\cite{Taylor:2015ppa}. The number of possible flips and flops grows as $h^{1,1}$ of the base grows. Hence to get the largest number of bases, the natural point of interest is the base with the largest $h^{1,1}$, which supports $X_4$ with the largest $h^{1,1}$ as well. Among the known set of Calabi-Yau fourfolds~\cite{Klemm:1996ts,Kreuzer:1997zg,Lynker:1998pb,Kreuzer:2001fu,Gray:2013mja,Scholler:2018apc}, such an $X_4$ has Hodge numbers
\be
(h^{1,1},h^{2,1},h^{3,1})=(303\,148,0,252)\,.
\ee
It was originally constructed as the Calabi-Yau hypersurface in the dual polytope of weighted projective space $\mb{P}^{1,1,84,516,1204,1816}$~\cite{Candelas:1997eh}, and the 4d F-theory gauge group on $X_4$ was read off by toric top methods:
\be
G=E_8^{2\,561}\times F_4^{7\,576}\times G_2^{20\,168}\times SU(2)^{30\,200}\,.
\ee
However, the smooth base threefold $B_3$ for $X_4$ with $h^{1,1}(B_3)=181\,819$ has not been constructed yet. Physically, it is interesting because the number of axions in the F-theory model on $X_4$ is the largest in the known 4d $\mc{N}=1$ string landscape~\cite{Grimm:2012yq}:
\be
\ba
N(\mathrm{axion})&=h^{1,1}(B_3)+1\cr
&=181\,820\,.
\ea
\ee

The construction of $B_3$ will be the main focus of this paper, which is discussed in section~\ref{sec:construction}, with the following steps:
\begin{enumerate}
\item{We start with a smooth non-compact toric threefold $B_{E_8}$ with 2561 rays and 5016 3d cones. On each of these 2561 toric divisors, we tune a Kodaira type $II^*$ singular fiber, which corresponds to $E_8$ gauge group in 4d F-theory. Then each of the 5016 3d cones give rise to an ``$E_8-E_8-E_8$ Yukawa point''~\cite{Apruzzi:2018oge}. There are also 7576 2d cones with 4d $(E_8,E_8)$ conformal matter.}
\item{Then we add two more rays into $B_{E_8}$ to make a smooth compact toric threefold $B_{\rm seed}$, which is called the ``seed'' of $B_3$.}
\item{We blow up each of the 5016 3d cones, such that the resulting base does not have any toric curves where the $(f,g)$ vanish to order to $(4,6)$ or higher, or any toric points where $(f,g)$ vanish to order $(8,12)$ or higher. We arrive at a toric base $B_{\rm toric}$ with $h^{1,1}(B_{\rm toric})=181\,200$ after this step.}
\item{Finally, it can be checked that for a generic fibration over $B_{\rm toric}$, there are 619 toric divisors with a non-Higgsable $E_8$ gauge group, such that $(f,g)$ vanish to order $(4,6)$ over a non-toric curve on each of these divisors. We then blow up these 619 non-toric curves and get the final $B_3$ with $h^{1,1}(B_3)=181\,819$.}
\end{enumerate}

In section~\ref{sec:4d}, we discuss some physical aspects of the 4d F-theory on $X_4$, including an argument for the ``saturation'' of the number of each geometric gauge group on the base $B_3$. We also estimated the number of self-dual $G_4$ flux choices: $10^{194\,000}$ on $X_4$. This number is much smaller than the number $10^{224\,000}$ on the elliptic Calabi-Yau fourfold with largest $h^{3,1}$\cite{Taylor:2015xtz}\footnote{Note that in~\cite{Taylor:2015xtz}, the number of $10^{272\,000}$ flux vacua is counted without applying the self-duality condition on $G_4$. After the self-duality condition is imposed, the number of flux vacua is reduced to $10^{224\,000}$.}.

In section~\ref{sec:flop}, we give a lower bound on the number of different smooth base threefolds that are related  by a number of flips and flops. This bound is proven to be $1100^{15\,048}\approx 7.5\times 10^{45\,766}$, which is much larger than any previous estimations in the literature. In particular, there exists a base that support a truly flat and smooth fibration $X_4$, see the local structure in figure~\ref{f:fullflat}. 

In section~\ref{sec:other}, we revisit the ``end point'' bases with $1000<h^{1,1}<13\,000$ studied in \cite{Taylor:2017yqr}. It is found that a number of these end point bases can be constructed from a seed  base with $E_8$ gauge groups in a similar way. We also explain the approximate ratio between the number of each gauge group and $h^{1,1}(B_3)$, which is observed in~\cite{Taylor:2017yqr}, as well as an approximate formula
\be
\frac{h^{1,1}(X_4)}{h^{1,1}(B_3)}\approx\frac{5}{3}
\ee
for the elliptic Calabi-Yau fourfolds with large $h^{1,1}$. The number of base flips and flops is lower bounded by the approximate formula
\be
N(\mathrm{flp})\gtrsim 10^{0.253\times h^{1,1}(B)}\,.
\ee

Finally, in section~\ref{sec:discussions}, we discuss the interpretation of large rank conformal matter coupled to gravity, as well as standard model building aspects.

\section{Construction of the maximal base}
\label{sec:construction}

In this section, we construct the non-toric base $B_3$ with $h^{1,1}(B_3)=181\,819$ that supports the elliptic Calabi-Yau fourfold $X_4$ with
\be
(h^{1,1},h^{2,1},h^{3,1})(X_4)=(303\,148,0,252)\,. 
\ee
For the discussions of toric threefold bases and 4d F-theory models over the base, we use the notations in the Section 2 of \cite{Taylor:2015ppa}. 



In \cite{Candelas:1997eh}, the elliptic Calabi-Yau fourfold $X_4$ is constructed as the anticanonical hypersurface of the dual  polytope of $\mathbb{P}^{1,1,84,516,1204,1806}$. Here we first perform an $SL(5,\mathbb{Z})$ rotation on $\mathbb{P}^{1,1,84,516,1204,1806}$, and get a polytope $\Delta_5$ with the following vertices:
\be
\ba
V(\Delta_5)=&\{(0,0,0,-1,1),(0,0,0,2,-1),(1,0,0,-1,-1),(0,1,0,-1,-1),\cr
&(0,0,1,-1,-1),(-1,-84,-516,-1,-1)\}\,.
\ea
\ee 

Its dual polytope $\Delta^\circ_5$ has the following vertices:
\be
\ba
V(\Delta^\circ_5)=&\{(0,0,0,1,0),(0,0,0,0,1),(-6,-6,1,-2,-3),(-6,37,-6,-2,-3),\cr
&(-6,-6,-6,-2,-3),(3606,-6,-6,-2,-3)\}\,,
\ea
\ee
which has a structure of a $\mb{P}^{2,3,1}$ bundle fibered over a 3d polytope $\Delta^\circ_3$ with vertices:
\be
V(\Delta^\circ_3)=\{(-6,-6,1),(-6,37,-6),(-6,-6,-6),(3606,-6,-6)\}\,.
\ee

To construct a smooth toric base threefold, we only select the subset $S\subset \Delta^\circ_3$ of lattice points $v_i=(v_{i,x},v_{i,y},v_{i,z})$ with 
\be
\mathrm{gcd}(v_{i,x},v_{i,y},v_{i,z})=1\,,
\ee
since if there exists $p>1$ with $p|v_{i,x},v_{i,y},v_{i,z}$, then any 3d cone containing $v_i$ has volume greater than one, which breaks the smoothness condition. In total, there are $|S|=181,203$ lattice points in this set, which correspond to the 1d rays of a compact toric threefold $B_{\rm toric}$ with
\be
h^{1,1}(B_{\rm toric})=181\,200\,.
\ee
In order to construct the list of 3d cones $\Sigma_3(B_{\rm toric})$ of $B_{\rm toric}$\footnote{Note that  $B_{\rm toric}$ is not weak-Fano, and the list of 3d cones is not a triangulation of the full polytope $\Delta^\circ_3$.}, we first pick the subset $S_{E_8}\subset S$ that corresponds to the divisors supporting Kodaira type $II^*$ singular fibers in $X_4$ (which carry $E_8$ geometric gauge groups in the 4d F-theory picture). To determine these rays, we consider the $\mc{F}$ and $\mc{G}$ polytope of $B_{\rm toric}$, defined as:
\be
\mc{F}=\{u\in\mb{Z}^3|\forall v_i\in V(\Delta^\circ_3)\ ,\ \langle u,v_i\rangle\geq -4\}\,,\label{mcF}
\ee
\be
\mc{G}=\{u\in\mb{Z}^3|\forall v_i\in V(\Delta^\circ_3)\ ,\ \langle u,v_i\rangle\geq -6\}\,.\label{mcG}
\ee

Especially, the polytope $\mc{G}$ has vertices:
\be
V(\mc{G})=\{(1,0,0),(0,1,0),(0,0,1),(-1,-84,-516)\}\,.
\ee

The order of vanishing of the Weierstrass polynomials $f\in\mc{O}(-4K_{B_{\rm toric}})$ and $g\in\mc{O}(-6K_{B_{\rm toric}})$ on $v_i$ are given by
\be
\mathrm{ord}_{v_i}(f)=\mathrm{min}_{u\in\mc{F}}(\langle u,v_i\rangle+4)\,,
\ee
\be
\mathrm{ord}_{v_i}(g)=\mathrm{min}_{u\in\mc{G}}(\langle u,v_i\rangle+6)\,.
\ee

There are in total 2561 $v_i\in S$ with $\mathrm{ord}_{v_i}(f)=4$, $\mathrm{ord}_{v_i}(g)=5$, which carries a type $II^*$ singular fiber and non-Higgsable $E_8$ gauge group~\cite{Morrison:2012np}. We define the set of these $v_i$ to be $S_{E_8}$. More explicitly, this set includes the following lattice points:
\be
\ba
S_{E_8}=&\{(-1,0,0),(1,0,0),(m,-1,0)\ (-1\leq m\leq 85)\ ,\cr
&(m,n,-1)\ (-1\leq n\leq 6,-1\leq m\leq 517-84n)\}\,.
\ea
\ee
Note that there does not exist a point in $S_{E_8}$ on the $z>0$ half-plane, hence the rays of $S_{E_8}$ only  form a non-compact toric threefold $B_{E_8}$. Nonetheless, there exists a triangulation of the convex hull of $S_{E_8}$, which gives rise to the set $\Sigma_3(B_{E_8})$ of 5016 3d cones with unit volume (not including the $z=0$ plane). The intersection of 3d cones in $\Sigma_3(B_{E_8})$ gives the set $\Sigma_2(B_{E_8})$ of 7576 2d cones. We present the detailed list of the lattice points and 3d cones in the supplementary Mathematica file~\cite{git:maxh11}. 

As $B_{E_8}$ is still non-compact, we add two more rays
\be
v_{2562}=(-6,-6,1)\ ,\ v_{2563}=(-6,37,-6)
\ee
into the toric fan of $B_{E_8}$. We also add the following 3d cones
\be
\ba
&\{\{(1,0,0),(-1,-1,0),(-6,-6,1)\},\{(m,-1,0),(m+1,-1,0),(-6,-6,1)\}\ (-1\leq m\leq 84),\cr
&\{(-1,0,0),(-6,-6,1),(-6,37,-6)\},\{(-1,0,0),(-1,6,-1),(-6,37,-6)\},\cr
&\{(m,6,-1),(m+1,6,-1),(-6,37,-6)\}\ (-1\leq m\leq 12),\{(13,6,-1),(1,0,0),(-6,37,-6)\}\cr
&\{(1,0,0),(-6,37,-6),(-6,-6,1)\}\}\,.\label{add-cones}
\ea
\ee
With these $2563$ rays and the 3d cones, we define a compact toric threefold denoted as the ``seed'' threefold $B_{\rm seed}$. Note that $B_{\rm seed}$ does not correspond to a reflexive polytope, and it is not weak Fano. Hence it can have a much larger $h^{1,1}$ than the ones discussed in~\cite{Halverson:2017ffz,Halverson:2017vde}. From the 4d F-theory perspective, we tune 2561 type $II^*$ singular fibers ($E_8$ gauge groups) on the rays in the set $S_{E_8}$. Then on the base $B_{\rm seed}$, there are 5016 toric points where $(f,g)$ vanishes to order equal or higher than $(8,12)$ ($(8,12)$-points) and 7576 toric curves where $(f,g)$ vanishes to order equal or higher than $(4,6)$ ($(4,6)$-curves). Thus the elliptic fibration is not flat, and we need to perform a sequence of base blow-ups. Note that the blow-up of $(4,6)$-curve or $(8,12)$-point does not change the number of complex structure moduli of the elliptic Calabi-Yau fourfold over it, because the set of Weierstrass monomials is unchanged after the blow  up. In \cite{Apruzzi:2018oge}, this type of base structure was called an ``$E_8-E_8-E_8$ Yukawa point'' along with 4d $(E_8,E_8)$ conformal matter, and the base blow-up sequence was already constructed. We will use such a blow-up procedure in this section.

\begin{figure}
\begin{center}
\includegraphics[width=12cm]{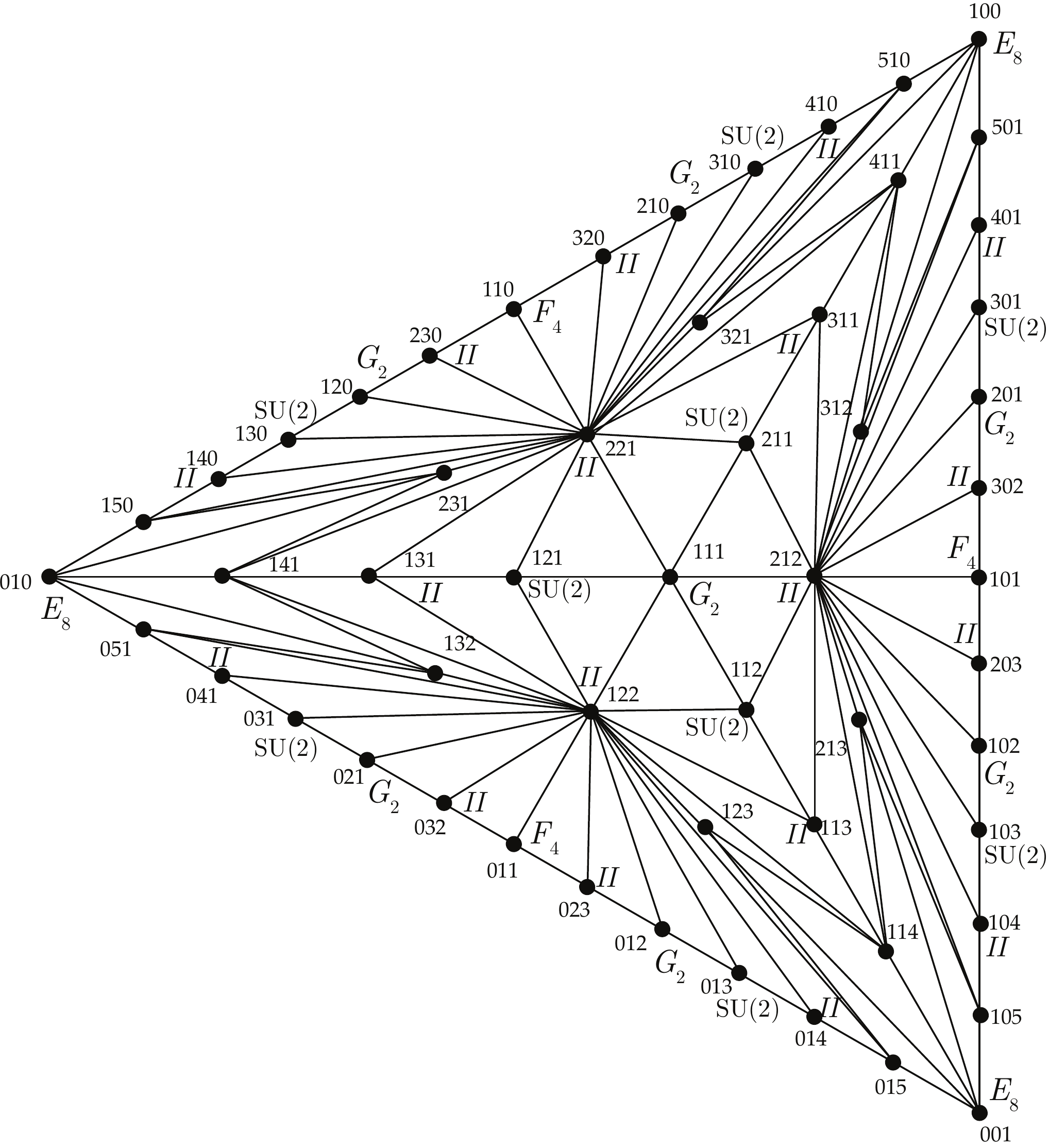}
\caption{The final 3d cones after blowing up the 3d cone $v_1 v_2 v_3$, where there are three $E_8$ geometric gauge groups on $v_i$. Each vertex $abc$ denotes an 1d ray $av_1+bv_2+cv_3$. It can be checked that all the 3d cones have unit volume if the original cone $v_1 v_2 v_3$ has unit volume. The geometric non-Higgsable gauge groups are also labelled on each vertex. We also label the Kodaira type II singular fiber on the divisors.}\label{f:fulltriangle}
\end{center}
\end{figure}

For each of the 5016 3d cones $\sigma_{i,3}=v_{i_1} v_{i_2} v_{i_3}\in\Sigma_3(B_{E_8})$, we denote the linear combination $av_{i_1}+bv_{i_2}+cv_{i_3}$ by $abc$. The blow-up of a toric point $v_{i_1} v_{i_2} v_{i_3}$ is then denoted by $(100,010,001;111)$, and the blow-up of a toric curve $v_{i_1} v_{i_2}$ is denoted by $(100,010;110)$. We perform the following sequences of blow-ups:
\be
\ba
Blp_1=&\{(100,010,001;111),(100,010,111;221),(100,001,111;212),\cr
&(010,001,111;122),(100,111;211),(010,111;121),(001,111;112),\cr
&(100,211;311),(010,121,131),(001,112,113)\cr
&(100,311;411),(010,131;141),(001,113;114)\},\label{Blp1}
\ea
\ee
\be
\ba
Blp_2=&\{(100,010;110),(100,110;210),(010,110;120),(110,210;320),\cr
&(110,120;230),(100,210;310),(100,310;410),(100,410;510),\cr
&(010,120;130),(010,130;140),(010,140;150),(100,001;101),\cr
&(100,101;201),(001,101;102),(101,201;302),(101,102;203),\cr
&(100,201;301),(100,301;401),(100,401;501),(001,102;103),\cr
&(001,103;104),(001,104;105),(010,001;011),(010,011;021),\cr
&(001,011;012),(011,021;032),(011,012;023),(010,021;031),\cr
&(010,031;041),(010,041;051),(001,012;013),(001,013;014),\cr
&(001,014;015)\},\label{Blp2}
\ea
\ee
\be
\ba
Blp_3=&\{(100,221;321),(100,212;312),(010,221;231),(010;122;132),\cr
&(001,212;213),(001,122;123)\}\,.\label{Blp3}
\ea
\ee
The final cones from blowing up $\sigma_3$ is shown in figure~\ref{f:fulltriangle}, which is the dual graph of figure. 8 in \cite{Apruzzi:2018oge}. There is no toric $(8,12)$ point or toric $(4,6)$ curve after the blow-up sequence.

The Kodaira singular fiber type and geometric non-Higgsable gauge groups on each toric divisor are given by (we also labelled the type II singular fiber, which does not have a non-Higgsable gauge group):
\be
\ba
&II^*\ ,\ E_8:100,010,001\cr
&IV^*_{ns}\ ,\ F_4:110,101,011\cr
&I^*_{0,ns}\ ,\ G_2:111,210,120,201,102,021,012\cr
&IV_{ns}\ ,\ SU(2):211,121,112,310,130,103,301,031,013\cr
&II:410,320,230,140,041,032,023,014,104,203,302,401,221,311,212,113,122,131\,.
\ea
\ee
The order of vanishing of $g$ over each divisor is given by the following table:
\be
\begin{tabular}{|c|c|c|c|c|c|c|}
\hline
\hline
& $\varnothing$ & $II$ & $SU(2)$ & $G_2$ & $F_4$ & $E_8$\\
\hline
$\mathrm{ord}(g)$ & 0 & 1 & 2 & 3 & 4 & 5\\
\hline
\end{tabular}
\ee

Note that the sequence of non-Higgsable gauge groups on each edge $v_1 v_2$, $v_1 v_3$, $v_2 v_3$ are exactly the same as the tensor branch of 6d minimal $(E_8,E_8)$ conformal matter~\cite{DelZotto:2014hpa}:
\be
E_8-\varnothing-II-SU(2)-G_2-II-F_4-II-G_2-SU(2)-II-\varnothing-E_8\,.\label{E8E8}
\ee

Finally, we get the smooth toric threefold $B_{\rm toric}$ after all the 5016 3d cones are blown up in this way. More precisely, if all the 5016 3d cones are blown up into figure~\ref{f:fulltriangle}, one needs to first perform the blow-ups (\ref{Blp1}) for all of these 3d  cones. Then one performs the blow-ups (\ref{Blp2}) for all the 7576 2d cones. Finally, one performs the blow-ups (\ref{Blp3}) for all the 5016 3d cones again. Note that the cones in (\ref{add-cones}) are also subdivided in the process. We provide the full list of $181\,203$ 1d rays and $362\,402$ 3d cones of $B_{\rm toric}$ in the supplementary Mathematica file~\cite{git:maxh11}.

The total numbers of non-Higgsable gauge groups are computed as follows. There is a single $F_4$ on each of the $7576$ edges on $B_{E_8}$, so the total number of $F_4$ is 7,576. For $G_2$, there are two of them on each of the $7576$ edges, and one in each of the 5016 3d cones of $B_{E_8}$, hence its total number is $20\,168$. For $SU(2)$, there are two of them on each of the $7576$ edges, and three of them in each of the 5016 3d cones of $B_{E_8}$, hence its total number is $30\,200$. 

On $B_{\rm toric}$, there are still a number of non-toric $(4,6)$-curves located on $E_8$ divisors, which needs to be blown up. They are analogous to the $(4,6)$-points on the $(-9)/(-10)/(-11)$ curves, in the elliptic Calabi-Yau threefold cases~\cite{Morrison:2012np,Morrison:2012js}. Given a ray $v_i$ with a non-Higgsable $E_8$ gauge group, we construct the set
\be
\mc{G}_5(v_i)=\{u\in\mb{Z}^3|\langle u,v_i\rangle+6=5\},\label{G5}
\ee
which corresponds to the monomials in the polynomial $g_5(v_i)$ of the expansion of the Weierstrass polynomial $g$:
\be
g=g_5(v_i) z_i^5+\mc{O}(z_i^6).
\ee
Here $z_i=0$ is the local hypersurface equation of the divisor corresponding to $v_i$. If there is more than one monomial in $g_5(v_i)$, then the equation
\be
g_5(v_i)=z_i=0
\ee
defines a $(4,6)$-curve on the base $B_{\rm toric}$, which needs to be blown up to get a fully flat elliptic fibration. On the other hand, if $g_5(v_i)$ only has a single monomial, then it has to be a constant complex number, as all the toric $(4,6)$-curves are already blown up. In total, there are 619 different $v_i$s with $|\mc{G}_5(v_i)|>1$, and we blow up the corresponding 619 $(4,6)$-curves (which are all irreducible). Note that the locations of these non-toric curves depend on the coefficients of $f$ and $g$, which correspond to the complex structure moduli of the elliptic Calabi-Yau fourfold.

After these non-toric blown ups, we get the non-toric base threefold $B_3$ with 
\be
\ba
h^{1,1}(B_3)&=h^{1,1}(B_{\rm toric})+619\cr
&=181\,819\,.
\ea
\ee
The generic elliptic fibration over $B_3$ has no codimension-two non-flat fiber. However, there are still codimension-three non-flat fiber over the points where $(f,g)$ vanish to order $(4,6)$ or higher. In figure~\ref{f:fulltriangle}, such points locate at the intersection of three divisors with $F_4(IV^*_{ns})-II-II$ and $G_2(I^*_{0,ns})-SU(2)-II$ singular fiber. Similar to the codimension-three non-flat fibers studied in the literature~\cite{Candelas:2000nc,Braun:2011ux,Braun:2013nqa,Bizet:2014uua,Baume:2015wia,Achmed-Zade:2018idx}, they will potentially lead to a tower of massless states and new Yukawa coupling terms. It is also notable that because of the $II-II$ collisions, the elliptic Calabi-Yau fourfold will have terminal singularities after the crepant resolution.

Nonetheless, in section~\ref{sec:flop}, we will show that there exists another configuration of 3d cones, such that the aforementioned loci are all absent. Thus it is possible to construct a smooth threefold base that supports a flat fibration $X_4$.

\section{4d F-theory on $X_4$}
\label{sec:4d}

\subsection{$X_4$ and the physical fields}

The 4d axions in the supergravity theories are given by the imaginary part of the K\"{a}hler moduli for the divisor classes on $B_3$, along with the reduction of 10d axiodilaton. The total number of axion fields is given by~\cite{Grimm:2012yq}:
\be
\ba
N(\mathrm{axion})&=h^{1,1}(B_3)+1\cr
&=181\,820\,.
\ea
\ee

As $X_4$ is a generic elliptic fibration over $B_3$, the geometric non-Higgsable gauge groups are already given in section~\ref{sec:construction}:
\be
G_{\rm nH}=E_8^{2\,561}\times F_4^{7\,576}\times G_2^{20\,168}\times SU(2)^{30\,200}\,.
\ee

The Hodge number $h^{1,1}(X_4)$ can be verified as:
\be
\ba
h^{1,1}(X_4)&=h^{1,1}(B_3)+\mathrm{rank}(G_{\rm nH})+1\cr
&=303\,148.
\ea
\ee

Finally, there are also a number of D3-branes in the 4d F-theory picture that can carry Abelian and non-Abelian gauge groups. The total number $N_{D_3}$ is bounded by the tadpole cancellation equation in the M-theory dual picture\footnote{We assume that the M-theory effective action can be approximated by the lowest order terms. It is possible that eight or higher derivative terms are significant, which breaks this assumption~\cite{Sethi:2017phn}.}:
\be
N_{D_3}+\frac{1}{2}\int_{X_4}G_4\wedge G_4=\frac{\chi(X_4)}{24}\,.
\ee
$\chi(X_4)$ is the Euler characteristic of $X_4$:
\be
\ba
\chi(X_4)&=6(8+h^{1,1}(X_4)+h^{3,1}(X_4)-h^{2,1}(X_4))\cr
&=1\,820\,448\,.
\ea
\ee

The self-dual $G_4$ flux satisfies
\be
\int_{X_4}G_4\wedge G_4\geq 0\,,
\ee
hence we have
\be
N_{D_3}\leq\frac{\chi(X_4)}{24}=75\,852\,.
\ee

\subsection{Saturation of gauge groups}
\label{sec:saturation}

On $B_3$, one cannot tune any larger non-Abelian gauge group on any toric divisor, because of the ``conformal matter'' structure in figure~\ref{f:fulltriangle}. Namely, the line (100,010), (100,001) and (010,001) has a sequence of gauge groups (\ref{E8E8}) that is the same as the tensor branch of 6d minimal $(E_8,E_8)$ conformal matter. The line (100,011), (010,101) and (001,110) also has the structure of 6d minimal $(E_8,F_4)$ conformal matter. Then any further tuning of gauge groups on a toric divisor will lead to additional toric $(4,6)$-curves on the base, which need to be blown up. For example, if we tune any non-Abelian gauge group on the divisor labelled by 131, then the curve (010,141) would be a $(4,6)$-curve. For the points $v_{2562}=(-6,-6,1)$ and $v_{2563}=(-6,37,-6)$ that are not in $B_{E_8}$, they also intersect divisors $D_i$ with $E_8$. Hence the presence of non-Abelian gauge group on these divisors would lead to toric $(4,6)$-curves $D_i\cdot D_{2562}$ and $D_i\cdot D_{2563}$ as well.

However, one can check that the base $B_3$ (and $B_{\rm toric}$) has an ``end point'' property (see \cite{Taylor:2017yqr}). Namely, any further toric blow-up leads to an invalid base with codimension-one locus with $\mathrm{ord}(f,g)\geq (4,6)$, for any fibration on the base. Hence we conclude that it is impossible to tune any larger non-Abelian gauge groups on the toric divisors. 

For the non-toric divisors on $B_3$, they either lie in the interior of the effective cone of $B_3$ or intersect a divisor with non-Higgsable $E_8$ gauge group. Thus it is expected that the tuning of any non-Abelian gauge group on these non-toric divisors would remove monomials in $f$ and $g$ and lead to a codimension-one $(4,6)$ locus as well. Similarly, if one tunes Abelian gauge groups generated by a non-trivial rational sections, it is also necessary to remove a number of monomials in $f$ and $g$~\cite{Morrison:2012ei,Wang:2016urs}. Hence we conjecture that any tuning of Abelian or non-Abelian gauge groups would lead to an invalid base with codimension-one $(4,6)$ loci.

Moreover, if one conjectures that the $h^{1,1}$ of elliptic Calabi-Yau fourfold (or even Calabi-Yau fourfold) is bounded by 303\,148, then any tuning of gauge groups on $B_3$ will exceed this bound. It is very interesting but difficult to prove this bound mathematically.

\subsection{Number of flux choices}

In \cite{Taylor:2015xtz}, the total number of $G_4$ flux choices on the Calabi-Yau fourfold $\mc{M}_{\rm max}$ with largest $h^{3,1}$ has been estimated following the logics of \cite{Ashok:2003gk,Douglas:2003um,Denef:2004ze,Denef:2008wq}. Namely, the different self-dual $G_4$ flux choices on a Calabi-Yau fourfold $X_4$ can be thought as lattice points in a sphere with radius $\sqrt{2Q}$ and dimension $b_4/2$, where
\be
Q=\frac{\chi(X_4)}{24}
\ee
and
\be
\ba
b_4&=2+2h^{3,1}+h^{2,2}\cr
&=4h^{1,1}-2h^{2,1}+6h^{3,1}+46
\ea
\ee
is the fourth Betti number of $X_4$.

For $\mc{M}_{\rm max}$ with
\be
(h^{1,1},h^{2,1},h^{3,1})=(252,0,303\,148)\,,
\ee
the Euler characteristic is $\chi=1\,820\,448$ and $b_4=1\,819\,942$. While for the $X_4$ with largest $h^{1,1}$, with
\be
(h^{1,1},h^{2,1},h^{3,1})=(303\,148,0,252)\,,
\ee
it has the same Euler characteristic as $\mc{M}_{\rm max}$, but a smaller $b_4=1\,214\,150$. Thus we expect that the total number of $G_4$ flux choices on $X_4$ is much smaller than that on $\mc{M}_{\rm max}$.

More explicitly, the counting of lattice points can be computed by \cite{mazo1990lattice}:
\be
N(b_4/2,Q)=\frac{1}{2\pi i}\int\frac{dt}{t}e^{-Qt}\vartheta_3(0,e^{t/2})^{b_4/2}\,,
\ee
where the integration goes from $i\infty$ to $-i\infty$, and $\vartheta_3$ is the Jacobi Theta function. We use a saddle point approximation $N(b_4/2,Q)\approx e^{S(t_*)}$, where $t_*$ is the critical point of the function
\be
S(t)=-\mathrm{ln}(-t)-Qt+\frac{b_4}{2}\mathrm{ln}\vartheta_3(0,e^{t/2})\,.
\ee
In our case, $b_4\approx 16Q$, hence the critical point is
\be
t_*\approx -3.63\,,
\ee
and we can estimate
\be
\ba
N(b_4/2,Q)&\approx 10^{2.56Q}
&\approx 10^{194\,000}\,.\label{M4flux}
\ea
\ee
This number is much smaller than the estimated self-dual flux choices on $\mc{M}_{\rm max}$, which is $10^{224\,000}$\cite{Taylor:2015xtz}.

Note that we have not taken into account the non-trivial metric on the space of self-dual $G_4$ flux\footnote{It was shown that in certain cases of CY3 and CY4, the volume of moduli space can be significantly smaller than the naive estimation, which leads to a smaller number of flux vacua~\cite{Cheng:2019mgz}.}, and we have not computed the number of flux vacua associated to each flux choice either. Again, if the higher derivative terms with $G_4$ in the M-theory effective action are included, the number is going to be corrected.

\section{Flip and flop phases}
\label{sec:flop}

\begin{figure}
\begin{center}
\includegraphics[width=9cm]{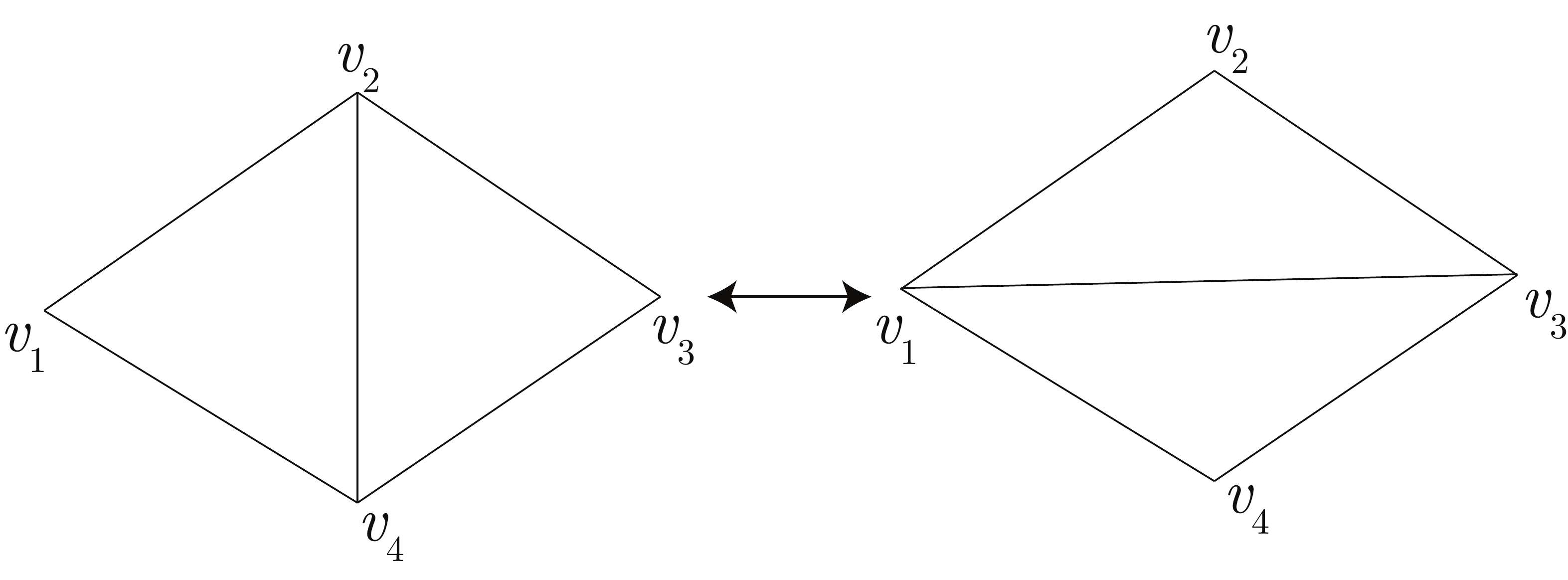}
\caption{The toric flip (flop) operation. For a flop, it is required that $v_1+v_3=v_2+v_4$.}\label{f:flop}
\end{center}
\end{figure}

In this section, we give a lower bound on the total number of topologically different smooth bases that are related to $B_3$ by a sequence of toric flips and flops. Such a local operation leads to a different set of 3d cones, while the set of 1d rays remains the same, see figure~\ref{f:flop}. In general, it is required that the four 1d rays involved satisfy $av_1+bv_3=cv_2+dv_4$, $(a,b,c,d\in\mb{Z})$. In particular, this operation is a flop if and only if $a=b=c=d=1$. In this section, we simply consider the flips and flops of the toric base $B_{\rm toric}$, and then perform the 619 blow-ups along the non-toric curves.

First, note that in figure~\ref{f:fulltriangle}, the divisors on the three $(E_8,E_8)$ edges are completely fixed. This means that we can consider the flips and flops inside such an $(E_8,E_8,E_8)$ triangle, and any set of 3d cones will be compatible with the other adjacent $(E_8,E_8,E_8)$ triangles (the philosophy is similar to the counting in~\cite{Halverson:2017ffz}). Moreover, we subdivide the triangle $(100,010,001)$ into six subsets: the three triangles $(100,010,221)$, $(010,001,122)$, $(100,001,212)$ and the three polygons $(100,221,010,111)$, $(100,111,001,212)$, $(010,111,001,122)$. We can individually evaluate the number of different sets of 3d cones in each of these subsets, and multiply these numbers together to get a lower bound on the total number of different base configurations within a single $(E_8,E_8,E_8)$ triangle.

\begin{figure}
\begin{center}
\includegraphics[width=14cm]{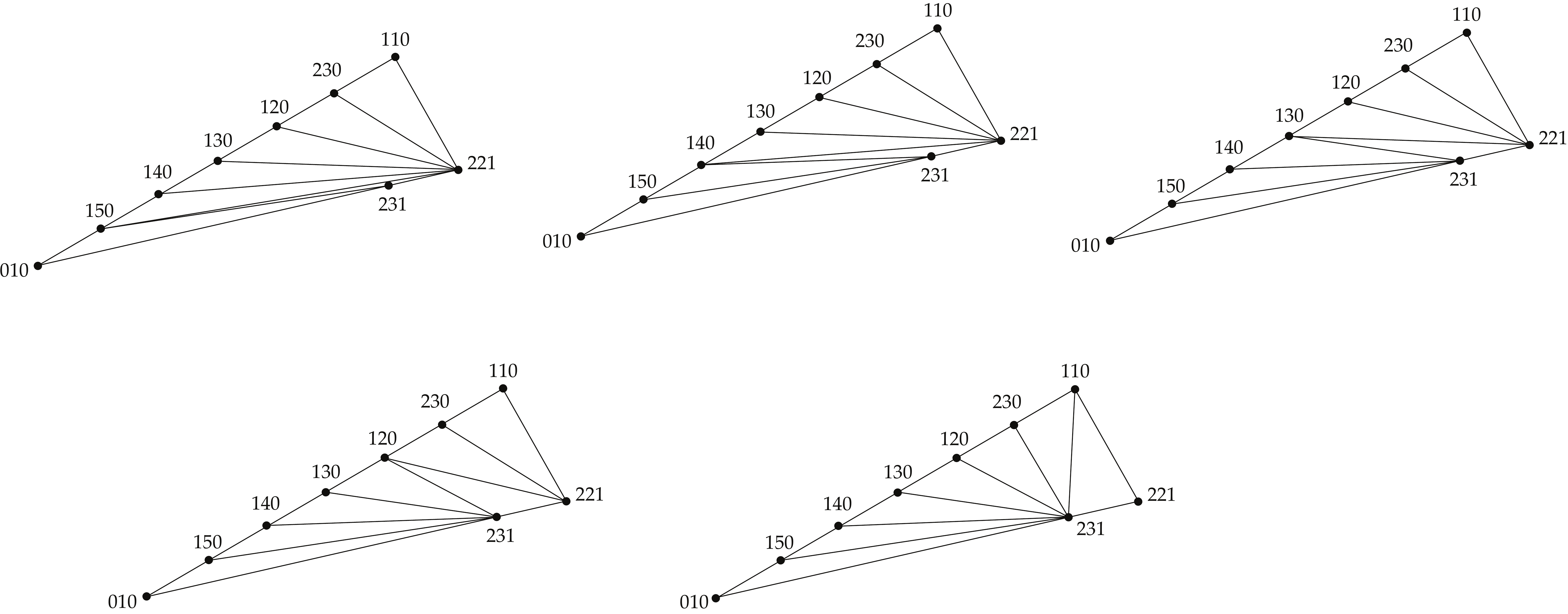}
\caption{All the possible sets of 3d cones of the triangle $(010,110,221)$, which is a subset of the full triangle in figure~\ref{f:fulltriangle}.}\label{f:triup}
\end{center}
\end{figure}

For the smaller triangle $(100,010,221)$, it can be further subdivided into two identical triangles $(010,110,221)$ and $(110,100,221)$. We plot all the five possible configurations of 3d cones of the triangle $(010,110,221)$ in figure~\ref{f:triup}. Note that the map from the bottom left to the bottom right configuration is a combination of two toric flips, instead of a simple toric flop. One can check that the volume of each 3d cone always equal to one, thus the base threefold is always smooth. Since there are in total six identical triangles of this shape in the full triangle $(100,010,001)$, there are $5^6$ different 3d cone configurations if the vertices $100$, $010$ and $001$ are taken to be inequivalent.

For the polygon $(100,221,010,111)$, we also plot 44 possible configurations of 3d cones in figure~\ref{f:flopdown}. One can again check that the volume of each 3d cone is always equals to one. As there are three identical polygons of this shape in the full triangle $(100,010,001)$, they contribute to $44^3$ different 3d cone configurations. Finally, the total number of 3d cone configurations within one of the $(E_8,E_8,E_8)$ triangle is lower bounded by
\be
\ba
N_{\rm flp,single}&=44^3\times 5^6\cr
&=1100^3.
\ea
\ee

\begin{figure}[!ht]
\begin{center}
\includegraphics[width=13cm]{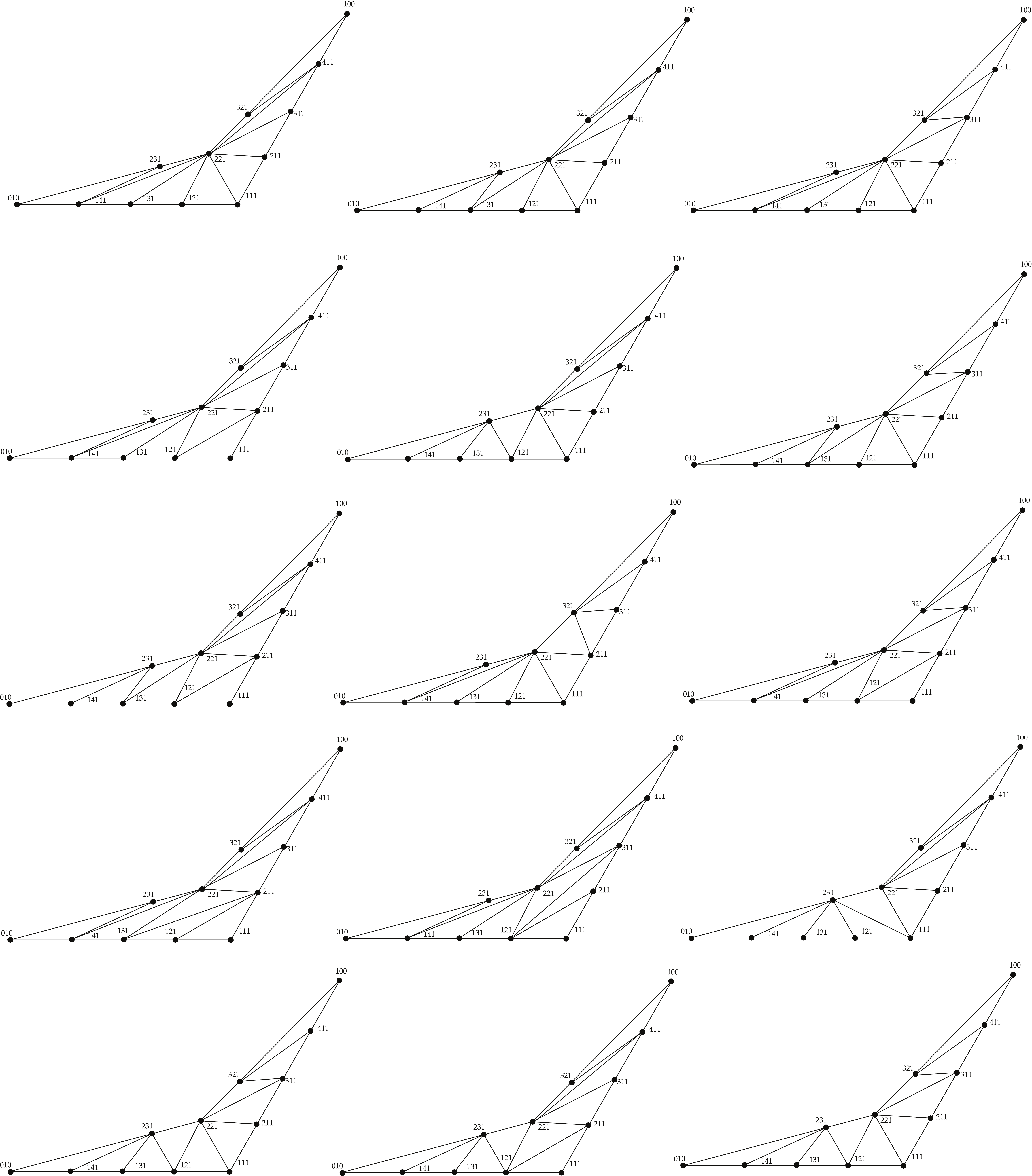}
\caption{The 44 sets of 3d cones of the polygon $(100,221,010,111)$, which is a subset of the full triangle in figure~\ref{f:fulltriangle}. }
\end{center}
\end{figure}

\begin{figure}[!ht]
\ContinuedFloat
\begin{center}
\includegraphics[width=13cm]{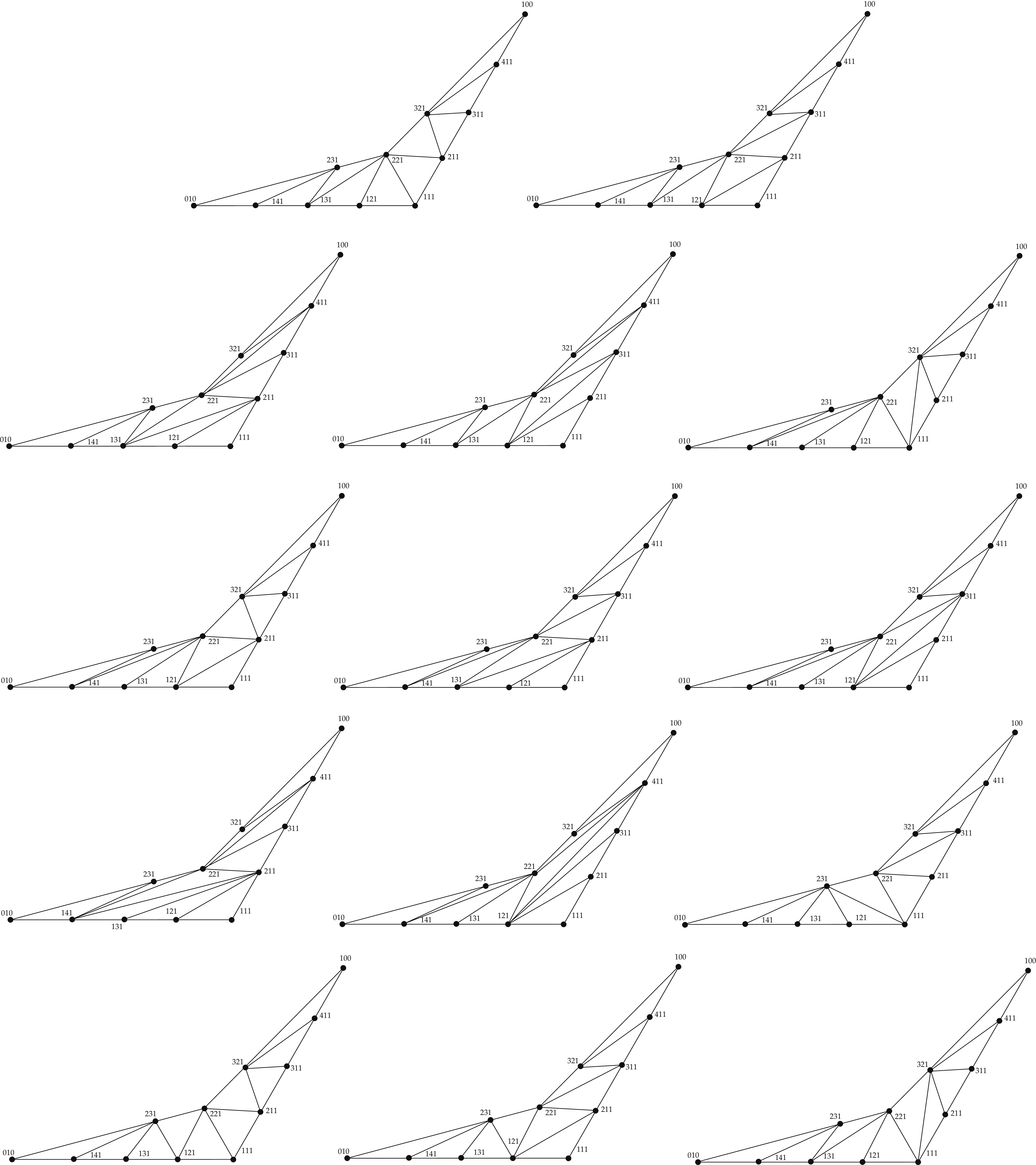}
\caption{The 44 sets of 3d cones of the polygon $(100,221,010,111)$, which is a subset of the full triangle in figure~\ref{f:fulltriangle}. \emph{(cont.)}}
\end{center}
\end{figure}

\begin{figure}
\ContinuedFloat
\begin{center}
\includegraphics[width=13cm]{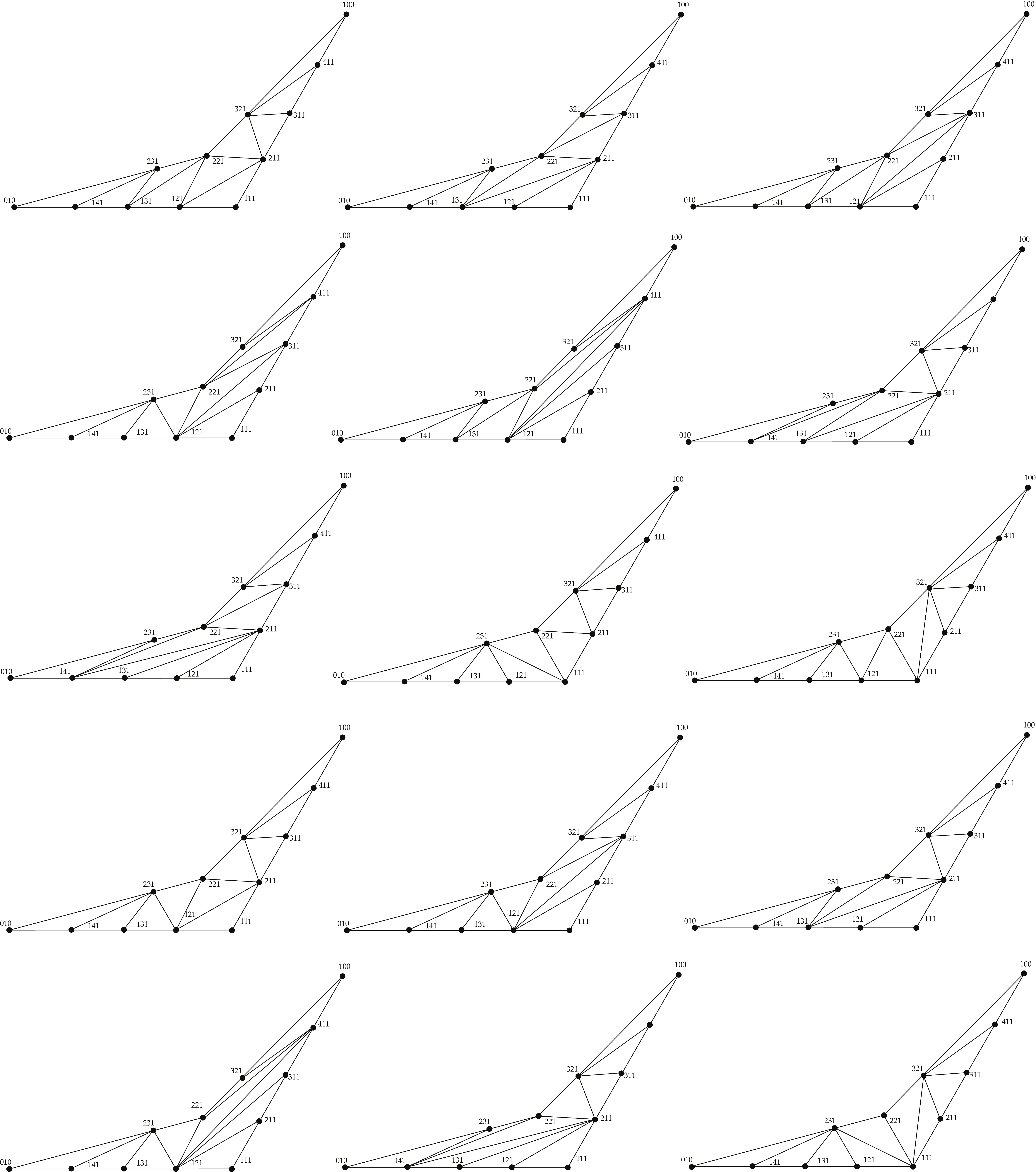}
\caption{The 44 sets of 3d cones of the polygon $(100,221,010,111)$, which is a subset of the full triangle in figure~\ref{f:fulltriangle}. \emph{(cont.)}}\label{f:flopdown}
\end{center}
\end{figure}

For the whole base $B_{\rm toric}$ with $5016$ $(E_8,E_8,E_8)$ triangles, every toric ray is considered as inequivalent since there is no reflexive toric automorphism. Thus the lower bound of the different flip and flop phases of $B_{\rm toric}$ is given by
\be
\ba
N_{\rm flp}(B_{\rm toric})&=N_{\rm flp,single}^{5\,016}\cr
&=1100^{15\,048}\cr
&\approx 7.5\times 10^{45\,766}\,.
\ea
\ee

It is amusing that after we multiply this number by the estimation of self-dual flux choices in (\ref{M4flux}), we get a number $\approx 10^{240\,000}$. It is bigger than the estimated number of self-dual flux choices on $\mc{M}_{\rm max}$, which is $10^{224\,000}$~\cite{Taylor:2015xtz}. We will briefly comment on the standard model building aspects of this model in section~\ref{sec:sm-building}.

\begin{figure}
\begin{center}
\includegraphics[width=12cm]{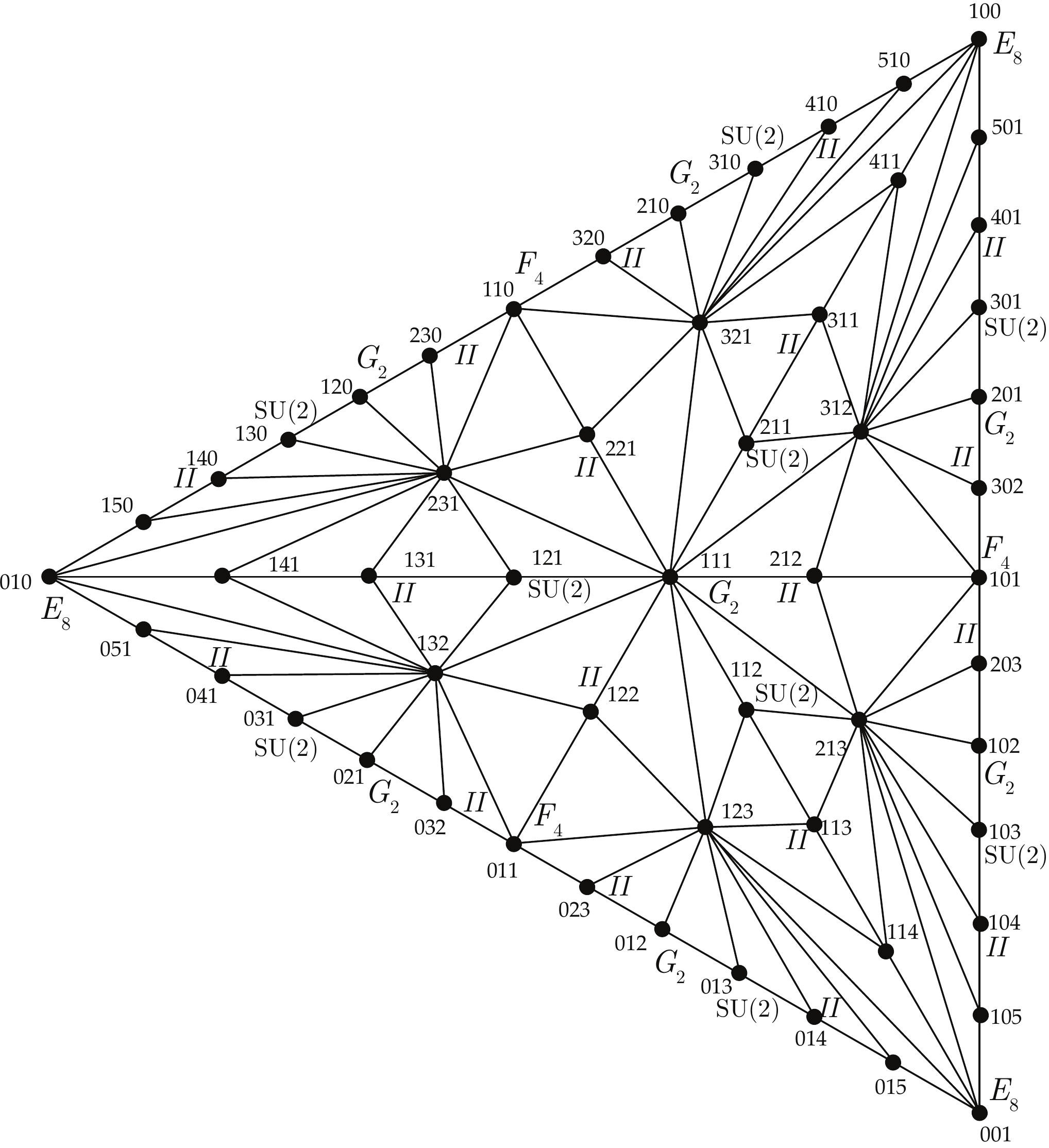}
\caption{The configuration of 3d cones in an $(E_8,E_8,E_8)$ triangle, such that all the codimension-three non-minimal loci and $II-II$ collision are absent. Each vertex $abc$ denotes an 1d ray $av_1+bv_2+cv_3$. It can be checked that all the 3d cones have unit volume if the original cone $v_1 v_2 v_3$ has unit volume. The geometric non-Higgsable gauge groups and the Kodaira type II singular fiber are labelled.}\label{f:fullflat}
\end{center}
\end{figure}

It is also notable that if one chooses the bottom right configuration of 3d cones in figure~\ref{f:triup} and figure~\ref{f:flopdown} for all these polygons, then the base supports a flat and smooth fibration $X_4$. One can check that all the codimension-three (4,6) loci and $II-II$ collisions are absent. We plot the subdivision of the $(E_8,E_8,E_8)$ triangle in figure~\ref{f:fullflat}. It is possible to directly generate this set of 3d cones from blowing up the triangle $(100,010,001)$:

\be
Blp_1=\{(100,010,001;111)\},\label{fBlp1}
\ee
\be
Blp_2=\{(100,010;110),(100,001;101),(010,001;011)\},\label{fBlp2}
\ee
\be
\ba
Blp_3=&\{(010,110,111;231),(100,110,111;321),(100,101,111;312),(001,101,111;213),\cr
&(001,011,111;123),(010,011,111;132),(110,111;221),(101,111;212),\cr
&(011,111;122),(010,111;121),(010,121;131),(010,131;141),(100,111;211),\cr
&(100,211;311),(100,311;411),(001,111;112),(001,112;113),(001,113;114)\},\label{fBlp3}
\ea
\ee
\be
\ba
Blp_4=&\{(100,110;210),(110,210;320),(100,210;310),(100,310;410),(100,410;510),\cr
&(010,110;120),(110,120;230),(010,120;130),(010,130;140),(010,140;150),\cr
&(100,101;201),(101,201;302),(100,201;301),(100,301;401),(100,401;501),\cr
&(001,101;102),(101,102;203),(001,102;103),(001,103;104),(001,104;105),\cr
&(010,011;021),(011,021;032),(010,021;031),(010,031;041),(010,041;051),\cr
&(001,011;012),(011,012;023),(001,012;013),(001,013;014),(001,014;015)\}.\label{fBlp4}
\ea
\ee
To construct the full $B_{\rm toric}$ from $B_{\rm seed}$, one needs to perform (\ref{fBlp1}) for all the $5016$ 3d cones. Then one perform (\ref{fBlp2}) for all the $7576$ 2d cones. Consequently, one performs (\ref{fBlp3}) for all the $5016$ 3d cones again. Finally, one performs (\ref{fBlp4}) for all the $7576$ 2d cones again.

\section{Other base threefolds}
\label{sec:other}

\subsection{End point bases and mirror pairs}

In \cite{Taylor:2017yqr}, the set of toric base threefolds is probed by random blow-up sequences from a starting point base, such as $\mb{P}^3$. The random blow-up sequences terminate at an ``end point'' base, where any further toric blow-up would lead to an invalid base with codimension-one (4,6) loci in the generic fibration. It is found that the $h^{1,1}$ of end point bases are concentrated at certain numbers. For example, about 10\% of the random blow-up sequences from $\mb{P}^3$ end up with toric bases with $h^{1,1}(B_{\rm toric})=1943$. 

In this paper, we show that a number of end point bases in \cite{Taylor:2017yqr} can be constructed by blowing up compact toric ``seed bases'' $B_{\rm seed}$ with tuned $E_8$ on the toric divisors. For any 3d cones with $E_8$ on all the three 1d rays, the structures of rays and cones can be chosen as figure~\ref{f:fulltriangle}. Then one can perform flips and flops to get more topologically distinct bases, as in section~\ref{sec:flop}.

For example, we can start with a weak Fano toric threefold $B_{\rm seed}$ with the following 29 rays:
\be
\ba
\{v_i\}=&\{(0, 0, 1), (0, 1, 0), (1, 0, 0), (-1, -1, -1), (1, 1, 1), (0, 1, 
  1),(-1, 0, 1), (2, 2, 1), \cr
&(0, 0, -1), (1, 0, 1), (1, -1, 0), (0, 1, -1), (0, -2, -1),(-1, 0, -1), (1, 2, 1), \cr
&(-2, -1, 
  0), (-3, -2, -1), (2, 0, 1), (-1, -1, 0), (0, -1, 0), (2, 1, 1), (1, 1, 0), \cr
&(-2, -2, -1), (0, -1, -1), (1, 2, 0), 
(-1, -2, -1), (-1, 0, 
  0), (-2, -1, -1), (2, 3, 1)\}\,.\label{P1P2rays}
\ea
\ee
We can choose the following set of 54 3d cones (where a number $i$ denotes the ray $v_i$):
\be
\ba
\{\sigma_3\}=&\{(13, 4, 26), (26, 19, 13), (10, 5, 18), (22, 25, 12), (25, 2, 
  12), (8, 5, 15), \cr
&(9, 3, 12), (12, 3, 22), (8, 25, 22), (20, 19, 
  1), (9, 4, 24), (27, 14, 2), \cr
&(10, 1, 5), (10, 18, 11), (18, 3, 
  11), (8, 22, 21), (21, 5, 8), (25, 29, 15),\cr
& (15, 29, 8), (8, 29, 
  25), (28, 17, 23), (28, 16, 17), (17, 16, 23), (28, 14, 16),\cr
& (4, 14,
   28), (13, 20, 11), (11, 24, 13), (16, 14, 27), (11, 1, 10), (5, 1, 
  6), \cr
&(11, 3, 24), (24, 3, 9), (20, 1, 11), (9, 14, 4), (22, 3, 
  21), (18, 5, 21),\cr
&(21, 3, 18), (2, 14, 12), (12, 14, 9), (15, 5, 
  6), (6, 2, 15), (15, 2, 25),\cr
& (24, 4, 13), (13, 19, 20), (19, 16, 
  7), (7, 16, 27), (19, 7, 1), (1, 7, 6),\cr
& (6, 7, 2), (2, 7, 27), (28, 
  23, 4), (4, 23, 26), (23, 16, 26), (26, 16, 19)\}\,.
\ea
\ee
One can check that the number of 2d cones is 81. After each of the 54 3d cones are blown up (for example according to figure~\ref{f:fulltriangle}), there are 19 new rays in the interior of each of 54 3d cones. On each of the 81 2d cones, there are 11 new rays. In total, we count the number of 1d rays in the blown up base $B_{\rm toric}$:
\be
N_1(B_{\rm toric})=29+54\times 19+81\times 11=1946.
\ee
Thus $h^{1,1}(B_{\rm toric})=1943$ exactly. The non-Higgsable gauge group on $B_{\rm toric}$ can be counted as:
\be
G_{\rm nH}=E_8^{29}\times F_4^{81}\times G_2^{216}\times SU(2)^{324}\,.
\ee

After the toric blow-ups, there are still 21 $E_8$ divisors with non-toric $(4,6)$-curves. Again this can be checked from the fact that the $\mc{G}_5$ polytopes (\ref{G5}) associated to these $E_8$ divisors have more than one lattice point. In this case, we can also check that these $(4,6)$-curves are all irreducible. From $B_{\rm toric}$ to $B_3$, one then needs to blow up these non-toric curves. Hence the total $h^{1,1}(X_4)$ is
\be
\ba
h^{1,1}(X_4)&=h^{1,1}(B_{\rm toric})+\mathrm{rank}(G_{\rm nH})+21+1\cr
&=3277\,.
\ea
\ee

It was also observed in \cite{Taylor:2017yqr} that this $X_4$ has exactly the mirror Hodge number of the generic elliptic CY4 over $\mb{P}^1\times\mb{P}^2$, which has $(h^{1,1},h^{3,1})=(3,3277)$\footnote{For the computations of Hodge numbers of elliptic CY4 over simple bases, also see~\cite{Klemm:1996ts,Mohri:1997uk}.}. Here we further observe that the convex hull of rays in (\ref{P1P2rays}) is exactly the dual polytope of $\mb{P}^1\times\mb{P}^2$, after an $SL(3,\mb{Z})$ rotation. A similar phenomenon has been observed in the case of elliptic CY3~\cite{Huang:2018vup}. In fact, the base for the elliptic CY3 with $(h^{1,1},h^{2,1})=(272,2)$ can be generated by blowing up a 2d seed base with the following nine rays:
\be
\{v_i\}=\{(1,0),(0,1),(-1,2),(-1,1),(-1,0),(-1,-1),(0,-1),(1,-1),(2,-1)\}\,.
\ee
Its dual polytope form the toric fan of a $\mb{P}^2$. After one tune nine $E_8$ gauge groups on these rays and blow up all the $(E_8,E_8)$ point into the full tensor branch, one gets the base in figure~2 of \cite{Huang:2018vup}. Then one needs to blow up the three $(-11)$-curves corresponding to rays $(-1,2)$, $(-1,-1)$ and $(2,-1)$ to get a non-toric base that supports a flat fibration.

Note that the seed base for $h^{1,1}(B_{\rm toric})=1943$ is not unique. One can also start with a base with the same number of rays
\be
\ba
\{v_i\}=&\{(0, 0, 1), (0, 1, 0), (1, 0, 0), (-1, -1, -1), (0, 1, 1), (1, 1, 
  1), (2, 1, 1), \cr
&(0, -1, -1), (-1, 0, 0), (-1, 1, 1), (2, 1, 
  2), (-1, -1, 0), (1, 1, 0), (-1, 0, -1), \cr
&(-1, -2, -1), (0, 1, 
  2), (2, 1, 0), (-1, 1, 2), (1, 1, 2), (1, 0, 1), (0, 0, -1), (0, 
  1, -1), \cr
&(-1, 0, 1), (-1, 1, -1), (1, 0, -1), (1, 1, -1), (-1, 1, 
  0), (2, 1, -1), (0, -1, 0)\}\label{1943rays2}
\ea
\ee
and 3d cones:
\be
\ba
\{\sigma_3\}=&\{(2, 10, 27), (10, 18, 23), (23, 18, 1), (6, 2, 13), (1, 19, 20), (13,
    28, 17), \cr
&(28, 3, 17), (13, 2, 26), (3, 29, 20), (20, 29, 1), (17, 
   6, 13), (1, 18, 16), \cr
&(16, 18, 5), (19, 11, 20), (19, 6, 11), (11, 
   7, 20), (11, 6, 7), (7, 3, 20),\cr
& (7, 6, 17), (17, 3, 7), (25, 28, 
   26), (25, 3, 28), (26, 28, 13), (9, 12, 4), \cr
&(5, 18, 10), (19, 5, 
   6), (6, 5, 2), (29, 12, 1), (1, 12, 23), (23, 12, 9),\cr
& (27, 10, 
   9), (14, 21, 22), (5, 10, 2), (27, 24, 2), (27, 9, 24), (9, 14, 
   24), \cr
&(24, 22, 2), (24, 14, 22), (10, 23, 9), (29, 15, 12), (12, 15,
    4), (4, 15, 8),\cr
& (8, 15, 29), (1, 16, 19), (19, 16, 5), (25, 8, 
   3), (25, 21, 8), (26, 21, 25), \cr
&(2, 22, 26), (22, 21, 26), (4, 14, 
   9), (4, 21, 14), (3, 8, 29), (8, 21, 4)\}\,.
\ea
\ee
After we tune $E_8$ on each of the 29 rays and perform the toric blow-ups, the resulting base is also an end point base with $h^{1,1}(B_{\rm toric})=1943$. Nonetheless, the convex hull of (\ref{1943rays2}) is different from (\ref{P1P2rays}). In fact, the dual polytope of (\ref{1943rays2}) has vertices
\be
\{v_i^\circ\}=\{(1,0,0),(0,-1,0),(-1,1,0),(0,0,1),(0,1,-1)\}\,,
\ee
which is the toric rays of a twisted $\mb{P}^2$ fibered over $\mb{P}^1$.

For the end points with $h^{1,1}(B_{\rm toric})=1727$, 2015, 2303 and 2591, we have checked that they can be generated by a simple seed base as well. Note that for $h^{1,1}(B_{\rm toric})=2591$, the seed base is exactly given by the maximal reflexive polytope in \cite{Halverson:2017ffz,Halverson:2017vde}. The details of the seed bases will be presented in appendix~\ref{app:seed}, and they form mirror pairs in a similar way.

\subsection{Seed bases with $F_4$}

For other end point bases in \cite{Taylor:2017yqr}, they may be generated in a similar way with different seed bases. In general, one can pick an end point base with large $h^{1,1}$, and select the 1d rays that carry non-Higgsable $E_8$. Such rays will naturally form the rays of the toric seed base, and the list of 3d cones can be generated by the triangulation of the convex hull of these rays. However, one need to check if the computed $N_1(B_{\rm toric})$ after the blow-ups matches the correct number of 1d ray. If they do not match, then the seed base needs to include more rays with other non-Higgsable gauge groups as well. 

For example, the class of 3d bases with $h^{1,1}(B_{\rm toric})=2249$ cannot be generated from a toric seed base with only $E_8$ non-Higgsable gauge groups. Nonetheless, we can start with a seed base $B_{\rm seed}$ with the following 34 rays:
\be
\ba
\{v_i\}=&\{(0, 0, 1), (0, 1, 0), (1, 0, 0), (-1, -1, -1), (0, 1, 1), (0, 
  0, -1), (1, 1, 1), \cr
&(-1, 0, -2), (-2, -1, -3), (1, 2, 3), (2, 4, 
  5), (0, -1, -1), (-2, -1, -4), (1, 2, 2), \cr
&(1, -1, -1), (1, 1, 2), (0, -1, 0), (2, 1, 2), (-1, -1, -2), (2, 3, 4), (0, -1, -2), \cr
&(1, 3, 3), (-1, -1, -3), (1, 0, 1), (0, 2, 1), (-3, -1, -5), (-2, 
  0, -3), (2, 0, 1), \cr
&(1, -1, 0), (2, -1, 0), (-1, 0, -1), (2, 2, 
  3), (-1, 1, -1), (-1, -2, 0)\}\label{2249rays}
\ea
\ee
and the 64 3d cones
\be
\ba
\{\sigma_3\}=&\{(14, 7, 18), (23, 13, 9), (27, 13, 23), (11, 20, 10), (11, 22, 
  20), (10, 22, 11), \cr
&(3, 30, 28), (28, 30, 24), (16, 28, 24), (15, 21,
   12), (15, 6, 21), (10, 32, 18), \cr
&(18, 32, 14), (10, 5, 25), (12, 29,
   15), (12, 4, 17), (10, 18, 16), (25, 5, 33), \cr
&(33, 2, 25), (15, 3, 
  6), (14, 32, 22), (24, 30, 29), (23, 8, 27), (10, 25, 22), \cr
&(22, 25, 
  14), (4, 9, 27), (1, 4, 31), (23, 9, 19), (3, 28, 7), (29, 30, 
  15),\cr
& (15, 30, 3), (10, 1, 5), (16, 1, 10), (22, 32, 20), (20, 32, 
  10), (27, 9, 26),\cr
& (9, 13, 26), (26, 13, 27), (17, 29, 12), (25, 2, 
  7), (31, 4, 27), (14, 25, 7), \cr
&(7, 28, 18), (18, 28, 16), (33, 8, 
  6), (5, 1, 33), (1, 31, 33), (6, 2, 33), \cr
&(24, 1, 16), (3, 2, 6), (7,
   2, 3), (17, 1, 29), (29, 1, 24), (21, 4, 12), \cr
&(31, 27, 33), (27, 8,
   33), (19, 9, 4), (23, 19, 21), (21, 8, 23), (6, 8, 21),\cr
& (21, 19, 
  4), (1, 4, 34), (1, 17, 34), (4, 17, 34)\}\,.\label{2249cones}
\ea
\ee

\begin{figure}
\begin{center}
\includegraphics[width=12cm]{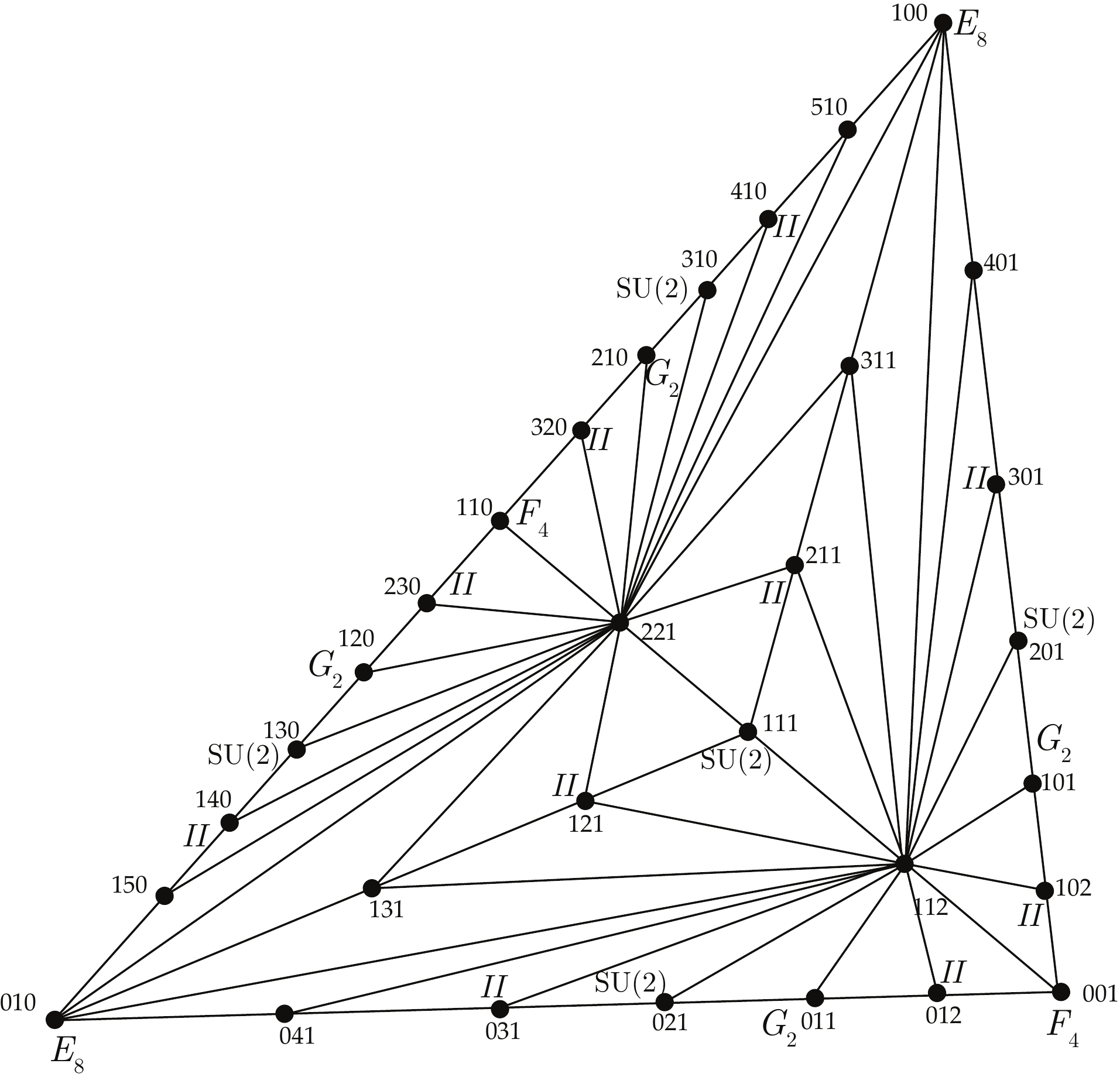}
\caption{The final 3d cones after blowing up the 3d cone $v_1 v_2 v_3$, where there are two $E_8$ geometric gauge groups on $v_1,v_2$ and an $F_4$ on $v_3$. Each vertex $abc$ denotes an 1d ray $av_1+bv_2+cv_3$. The geometric non-Higgsable gauge groups are also labelled on each vertex. We also label the Kodaira type II singular fiber on the divisors.}\label{f:E8E8F4}
\end{center}
\end{figure}

Now we tune 33 $E_8$ gauge groups on the rays $v_1,\dots,v_{33}$ and an $F_4$ gauge group on $v_{34}$. Then the first 61 3d cones of (\ref{2249cones}) can be blown up according to figure~\ref{f:fulltriangle}. But the last three of the 3d cones correspond to an $(E_8,E_8,F_4)$ collision, which can be blown up to  figure~\ref{f:E8E8F4}. We denote the two rays with $E_8$ by $v_1,v_2$ and the ray with $F_4$ by $v_3$. We use $abc$ to denote a ray $av_1+bv_2+cv_3$. Then the blow-up sequence is:

\be
\ba
Blp=&\{(100,010,001;111),(100,010,111;221),(111,001;112),(100,010;110),\cr
&(100,110;210),(010,110;120),(110,210;320),(110,120;230),\cr
&(100,210;310),(100,310;410),(100,410;510),(010,120;130),\cr
&(010,130;140),(010,140;150),(100,111;211),(100,211;311),\cr
&(010,111;121),(010,121;131),(100,001;101),(100,101;201),\cr
&(100,201;301),(100,301;401),(101,001;102),(010,001;011),\cr
&(010,011;021),(010,021;031),(010,031;041),(001,011;012)\}\,.
\ea
\ee

Now we can count the total number of $B_{\rm toric}$ after the blow-ups: there are 34 rays in (\ref{2249rays}), 19 rays in the interior of each of the 61 $(E_8,E_8,E_8)$ 3d cones, 7 rays in the interior of each of the 3 $(E_8,E_8,F_4)$ 3d cones, 11 rays in the interior of each of the 93 $(E_8,E_8)$ 2d cones and 5 rays in the interior of each of the 3 $(E_8,F_4)$ 2d cones. In total, there are exactly 2252 rays in $B_{\rm toric}$, which gives rise to the correct $h^{1,1}(B_{\rm toric})=2249$!

The non-Higgsable gauge groups on $B_{\rm toric}$ can be counted:
\be
G_{\rm nH}=E_8^{33}\times F_4^{94}\times G_2^{250}\times SU(2)^{375}\,.
\ee

After the toric blow-ups, there are still 21 $E_8$ divisors with non-toric $(4,6)$-curve. From $B_{\rm toric}$ to $B_3$, one needs to blow up these non-toric curves. Hence the total $h^{1,1}(X_4)$ is
\be
\ba
h^{1,1}(X_4)&=h^{1,1}(B_{\rm toric})+\mathrm{rank}(G_{\rm nH})+21+1\cr
&=3786\,.
\ea
\ee

\subsection{Estimated number of gauge factors}

Here we provide an explanation of the gauge group counting formula (4.1) in \cite{Taylor:2017yqr}. In a seed base with a large number $N$ of 1d rays, the number of 2d and 3d cones generally scale as:
\be
\ba
N_2(B_{\rm seed})&\approx 3N\cr
N_3(B_{\rm seed})&\approx 2N\,.
\ea
\ee
Assuming that each 1d ray has a geometric $E_8$ gauge group, then after the seed base is fully blown up to the base $B_{\rm toric}$ with no toric codimension-two (4,6) locus, the total number of 1d rays is given by
\be
\ba
N_1(B_{\rm toric})&\approx N+19\times 2N+11\times 3N\cr
&\approx 72N\,.
\ea
\ee
The number $N_1(B_{\rm toric})\approx h^{1,1}(B_{\rm toric})$ if the number is large. The total numbers of geometric non-Higgsable gauge groups of each type are given by:
\be
\ba
N(E_8)&=N\cr
N(F_4)&=N_2(B_{\rm seed})\approx 3N\cr
N(G_2)&=2N_2(B_{\rm seed})+N_3(B_{\rm seed})\approx 8N\cr
N(SU(2))&=2N_2(B_{\rm seed})+3N_3(B_{\rm seed})\approx 12N\,.
\ea
\ee
Thus we have the approximate formula for the number of each type of gauge groups 
\be
\ba
N(E_8)&\approx\frac{h^{1,1}(B_{\rm toric})}{72}\cr
N(F_4)&\approx\frac{h^{1,1}(B_{\rm toric})}{24}\cr
N(G_2)&\approx\frac{h^{1,1}(B_{\rm toric})}{9}\cr
N(SU(2))&\approx\frac{h^{1,1}(B_{\rm toric})}{6}\,.\label{N-gaugegroup}
\ea
\ee

We can also estimate the ratio between $h^{1,1}(B_3)$ and $h^{1,1}(X_4)$ based on (\ref{N-gaugegroup}) and
\be
h^{1,1}(X_4)=h^{1,1}(B_3)+\mathrm{rank}(G_{\rm nH})+1\,.
\ee
We use an approximation $h^{1,1}(B_3)\approx h^{1,1}(B_{\rm toric})$, although in principle the $B_3$ base is generated by blowing up $B_{\rm toric}$ along non-toric curves. Then we arrive at the following approximate formula for the end point bases with large $h^{1,1}(B_3)$:
\be
\frac{h^{1,1}(X_4)}{h^{1,1}(B_3)}\approx\frac{5}{3}\,.\label{h11ratio}
\ee

For the number of flip and flop phases on $B_{\rm toric}$, the lower bound can also be estimated with the methods in section~\ref{sec:flop}:
\be
\ba
N(\mathrm{flp})&\gtrsim (1100)^{3N_3}\cr
&\approx (1100)^{h^{1,1}(B_{\rm toric})/12}\cr
&\approx 10^{0.253\times h^{1,1}(B_{\rm toric})}\,.\label{flop-estimation}
\ea
\ee
Note that if there exists toric automorphism on the seed base $B_{\rm seed}$, then this number will be reduced by $\mc{O}(10^1)$, which has no significant change on the exponential. 

For example, for the number of $B_{\rm toric}$ with $h^{1,1}(B_{\rm toric})=1943$, it is estimated to be 
\be
N(\mathrm{flp})\gtrsim 10^{493}.
\ee

This number is much bigger than the estimated number of ``good bases'' (the bases without codimension-two (4,6) locus) with the same $h^{1,1}$ in~\cite{Taylor:2017yqr}, which is around $10^{200}$. This suggests that the statistical methods in~\cite{Taylor:2017yqr} lead to a systematic  underestimation.

\section{Discussions}
\label{sec:discussions}

\subsection{Supergravity coupled to conformal matter}

From the construction of base threefolds with large $h^{1,1}$, we see that the $(E_8,E_8,E_8)$ collision is prevalent. In fact, the Calabi-Yau with largest known $h^{1,1}$ provides the known example of the highest rank conformal matter coupled to supergravity. For 6d (1,0) theories, the elliptic Calabi-Yau threefold $X_3$ with the largest $h^{1,1}(X_3)=491$ has the following toric base geometry~\cite{Taylor:2012dr}:
\be
(-12//-11//(-12//)^{13},-11//-12,0)\,.
\ee
Each number denotes the self-intersection number of each $\mb{P}^1$ curve on the base, which intersects each other in a cyclic way. The symbol ``//'' denotes the following chain of curves in the tensor branch of minimal $(E_8,E_8)$ conformal matter:
\be
//\equiv -1,-2,-2,-3,-1,-5,-1,-3,-2,-2,-1\,.
\ee
To get a base without $(4,6)$-points, the two $(-11)$-curves need to be blown up at a non-toric point as well. One can see that the non-minimal $(E_8,E_8)$ conformal matter\cite{DelZotto:2014hpa,Ohmori:2015pia} with order $N=16$ can be embedded into this base. Such a 6d (1,0) SCFT has the following tensor branch in the standard notation:
\be
[E_8]-1-2-\overset{\mathfrak{su}(2)}{2}-\overset{\mathfrak{g}_2}{3}-1-\overset{\mathfrak{f}_4}{5}-1-\overset{\mathfrak{g}_2}{3}-\overset{\mathfrak{su}(2)}{2}-2-1-\overset{\mathfrak{e}_8}{12}-1-\dots-1-[E_8]\,,
\ee
where there are 15 $(-12)$-curves with $\mathfrak{e}_8$ gauge group in the middle.

For 5d $\mc{N}=1$ theories, similarly one can consider M-theory on the resolved compact Calabi-Yau threefold $X_3$. Then we can couple 5d supergravity with the KK reduction of the non-minimal $(E_8,E_8)$ conformal matter with $N=16$, which has the following 5d IR quiver gauge theory description with rank $r=471$~\cite{Ohmori:2015pia}:
\be
\begin{array}{ccc}
& SU(48) &\\
& \vert & \\
SU(16)-SU(32)-SU(48)-SU(64)-SU(80)-&SU(96)&-SU(64)-SU(32)\,.
\end{array}
\label{maxquiver}
\ee
However, to get a genuine 5d SCFT fixed point, one needs to decouple an $SU(16)$ vector multiplet from the theory, which geometrically corresponds to decompactifying 15 divisors in $X_3$~\cite{Apruzzi:2019kgb,Apruzzi:2019opn}. After the decompactification, the gravity sector will be decoupled again. Nonetheless, the 5d quiver (\ref{maxquiver}) is still the quiver gauge theory with the known largest rank that can be coupled to 5d $\mc{N}=1$ supergravity.

Finally, for the 4d $\mc{N}=1$ theories, it is unclear whether the $E_8-E_8-E_8$ Yukawa point  actually corresponds to an SCFT fixed point or not~\cite{Apruzzi:2018oge}. Nonetheless, the $X_4$ with largest $h^{1,1}(X_4)=303\,148$ in this paper provides the example of 4d supergravity coupled to an $E_8$ quiver network with the largest known number (2561) of $E_8$ gauge groups.

It would be fascinating to have a swampland bound argument along the philosophy of \cite{Heckman:2019bzm,Kim:2019vuc,Lee:2019skh,Kim:2019ths}, for the various cases discussed here: 6d (1,0), 5d $\mc{N}=1$ and 4d $\mc{N}=1$. Alternatively, one can also attempt to challenge these bounds in other parts of the  string landscape with a supergravity sector.

\subsection{Standard model building}
\label{sec:sm-building}

Another interesting question is whether the 4d F-theory model on $X_4$ with the largest $h^{1,1}(X_4)$ has any model building implications. Since this model has the largest number of axions 
\be
N(\mathrm{axions})=181\,820
\ee
in the known 4d superstring landscape, it is potentially useful for the inflation models with a large number of axions, e. g. \cite{Liddle:1998jc,Dimopoulos:2005ac,Easther:2005zr,Grimm:2007hs}.

Nonetheless, it is hard to realize the standard model gauge group on $X_4$ geometrically, because of a similar problem to the one discussed in \cite{Taylor:2015xtz,Tian:2018icz}. If we embed $G_{\rm sm}=SU(3)\times SU(2)\times U(1)$ into a single $E_8$, then it is implausible to get chiral families~\cite{Tatar:2006dc}. On the other hand, if we embed $G_{\rm sm}=SU(3)\times SU(2)\times U(1)$ into a single $F_4$, the branching rule does not give the correct standard model hypercharges. Moreover, it is impossible to tune any larger gauge groups on the base, if we accept that $h^{1,1}(X_4)$ is already maximal (also see the discussions in section~\ref{sec:saturation}).

Thus it is more plausible to realize a part or all of the $G_{\rm sm}$ as the gauge bosons from D3 branes. Then the geometric gauge groups will be treated as dark matter sectors, if they do not intersect these D3 branes.

\subsection*{Acknowledgements}

The author thanks Fabio Apruzzi, James Halverson, Ben Heidenreich, Cody Long, Liam McAllister, Tom Rudelius, Sakura Schafer-Nameki, Jiahua Tian, Washington Taylor and Dan Xie for discussions. 
This work is supported by the  ERC Consolidator Grant number 682608 ``Higgs bundles: Supersymmetric Gauge Theories and Geometry (HIGGSBNDL)''. 

\appendix

\section{Other toric seed threefolds}
\label{app:seed}

In section~\ref{sec:other}, we have presented some toric seed threefolds for the end point bases in \cite{Taylor:2017yqr} with $h^{1,1}(B_{\rm toric})\sim\mc{O}(10^3)$. In this section, we will present four more toric seed threefolds with $E_8$ on each of the toric rays.

\begin{enumerate}
\item{Toric seed threefold $B_{\rm seed}$ for $h^{1,1}(B_{\rm toric})=1727$

There are 26 1d rays in $B_{\rm seed}$:
\be
\ba
\{v_i\}=&\{(0,0,1), (1,0,0), (0,0,-1), (1,1,0), (1,1,2), (-1,0,1), (2,1,-1),\cr
& (-1,-1,0), (-1,0,0), (-1,0,-2), (1,1,-1), (-2,-1,-1), (2,1,0), (3,1,0),\cr
& (-3,-1,-1), (-1,0,2), (1,1,1), (-3,-1,-2), (-3,-1,0), (1,1,-2), (-1,0,-1),\cr
& (-2,-1,0), (2,1,1), (-3,-1,1), (-2,-1,1), (-3,-1,2)\}\label{1727rays}
\ea
\ee
and 48 3d cones:
\be
\ba
\{\sigma_3\}=&\{  (13, 4, 7),  (13, 7, 14),  (7, 2, 14),  (4, 23, 17),  (22, 19, 25),  (13, 
  23, 4), \cr
& (19, 9, 6),  (6, 9, 4),  (12, 15, 19),  (19, 15, 9),  (14, 23, 
  13),  (14, 2, 23), \cr
& (5, 16, 6),  (1, 16, 5),  (25, 8, 22),  (9, 15, 
  21),  (17, 23, 5),  (23, 1, 5),\cr
&  (5, 6, 17),  (4, 21, 11),  (18, 12, 
  10),  (18, 15, 12),  (10, 15, 18),  (2, 1, 23), \cr
& (21, 15, 10),  (7, 4, 
  11),  (11, 20, 7),  (20, 2, 7),  (3, 2, 20),  (17, 6, 4), \cr
& (24, 19, 
  6),  (25, 19, 24),  (25, 24, 26),  (24, 6, 26),  (25, 16, 1),  (1, 8, 
  25),  \cr
&(6, 16, 26),  (26, 16, 25),  (2, 8, 1),  (3, 8, 2),  (4, 9, 
  21),  (8, 12, 22), \cr
& (20, 10, 3),  (11, 10, 20),  (21, 10, 11),  (3, 12, 
  8),  (10, 12, 3),  (22, 12, 19)\}\,.
\ea
\ee

After the full blow-up to $h^{1,1}(B_{\rm toric})$, the non-Higgsable gauge group is
\be
G_{\rm nH}=E_8^{26}\times F_4^{72}\times G_2^{192}\times SU(2)^{288}\,.
\ee
There are 19 $E_8$ divisors with a non-toric $(4,6)$-curve. Among them, the non-toric $(4,6)$-curve on the ray $(1,0,0)$ has two irreducible components. Hence from $B_{\rm toric}$ to $B_3$, one needs to blow up these 20 non-toric curves, and the total $h^{1,1}(X_4)$ is
\be
\ba
h^{1,1}(X_4)&=h^{1,1}(B_{\rm toric})+\mathrm{rank}(G_{\rm nH})+20+1\cr
&=2916\,.\label{2916h11}
\ea
\ee
The dual polytope of (\ref{1727rays}) form a toric fan with rays $\{(0,-1,0)$, $(0,1,0)$, $(1,-2,0)$, $(-1,2,-1)$, $(-1,2,1)$, $(-1,2,0)\}$. The generic elliptic CY4 on this toric threefold has $(h^{1,1},h^{3,1})=(4,2916)$, which should be the mirror of (\ref{2916h11}).

}

\item{Toric seed threefold $B_{\rm seed}$ for $h^{1,1}(B_{\rm toric})=2015$

There are 30 1d rays in $B_{\rm seed}$:
\be
\ba
\{v_i\}=&\{(0, 0, 1), (0, 1, 0), (1, 0, 0), (-1, -1, -1), (0, 1, 
  1), (0, -1, -1), (0, 1, 2),\cr
& (0, -1, 0), (0, 0, -1), (-1, 
  0, -1), (-1, 1, -1), (-1, -1, 0), (-1, -1, 1), (-1, 0, 0),\cr
& (-1, 0, 
  1), (-1, 1, 2), (-1, -1, -2), (-1, -2, -2), (1, 1, 1), (-1, 1, 
  0), (-1, 1, 3), \cr
&(-1, 0, -2), (-1, -2, -3), (-1, -2, -1), (-1, 0, 
  2), (-1, -2, -4), (-1, -2, 0), (-1, 1, 
  1), \cr
&(-1, -1, -3), (0, -1, -2)\}\label{2015rays}
\ea
\ee
and 56 3d cones:
\be
\ba
\{\sigma_3\}=&\{(29, 17, 23), (2, 20, 11), (11, 20, 10), (30, 22, 29), (22, 17, 
  29), (29, 26, 30),\cr
& (29, 23, 26), (26, 23, 30), (9, 30, 3), (3, 30, 
  6), (8, 13, 1), (12, 13, 27),\cr
& (27, 13, 8), (7, 5, 19), (19, 1, 
  7), (5, 2, 19), (19, 2, 3), (3, 1, 19), \cr
&(8, 3, 6), (27, 24, 4), (6, 
  24, 27), (2, 11, 9), (5, 28, 2), (7, 16, 5),\cr
& (17, 18, 23), (23, 18, 
  30), (8, 1, 3), (3, 2, 9), (4, 12, 27), (1, 25, 7), \cr
&(14, 15, 
  12), (28, 16, 15), (9, 22, 30), (30, 18, 6), (22, 11, 10), (9, 11, 
  22), \cr
&(16, 21, 25), (25, 21, 7), (7, 21, 16), (6, 27, 8), (22, 10, 
  4), (4, 18, 17),\cr
& (12, 15, 13), (1, 13, 25), (13, 15, 25), (5, 16, 
  28), (28, 15, 14), (4, 17, 22),\cr
& (15, 16, 25), (10, 20, 14), (4, 14, 
  12), (10, 14, 4), (6, 18, 24), (24, 18, 4),\cr
& (28, 20, 2), (14, 20, 
  28)\}\,.
\ea
\ee

After the full blow-up to $h^{1,1}(B_{\rm toric})$, the non-Higgsable gauge group is
\be
G_{\rm nH}=E_8^{30}\times F_4^{84}\times G_2^{224}\times SU(2)^{336}\,.
\ee
There are 20 $E_8$ divisors with non-toric $(4,6)$-curves. For the single divisor corresponding to the ray $(-1,-2,0)$, the $(4,6)$-curve has two irreducible components. Hence there are in total 21 non-toric curves to be blown up in order to get $B_3$. Hence the total $h^{1,1}(X_4)$ is
\be
\ba
h^{1,1}(X_4)&=h^{1,1}(B_{\rm toric})+\mathrm{rank}(G_{\rm nH})+21+1\cr
&=3397\,.
\ea
\ee

The dual polytope of (\ref{2015rays}) form the toric fan of generalized Hirzebruch surface $\tilde{\mb{F}}_1$. The generic elliptic fibration over $\tilde{\mb{F}}_1$ exactly has Hodge numbers $(h^{1,1},h^{3,1})=(3,3397)$.

}

\item{Toric seed threefold $B_{\rm seed}$ for $h^{1,1}(B_{\rm toric})=2303$

There are 34 1d rays in $B_{\rm seed}$:
\be
\ba
\{v_i\}=&\{(0, 0, 1), (0, 1, 0), (1, 0, 0), (-1, -1, -1), (1, 1, 1), (-1, 
  0, -1), (0, 1, 1),\cr
& (0, 1, -1), (0, -1, 0), (1, 2, 
  2), (0, -1, -1), (1, -1, 1), (1, -1, 0), (1, 2, 3),\cr
& (0, 
  0, -1), (-1, -1, -2), (1, 2, 1), (1, 1, 0), (1, 1, 
  2), (-1, -2, -2), (1, 2, 0),\cr
& (-2, -1, -3), (1, -2, 
  0), (0, -2, -1), (1, 0, 1), (1, 0, 2), (0, 1, 2), (-2, -1, -2),\cr
& (-1,
   0, -2), (-3, -2, -4), (-1, 0, 0), (1, 1, 3), (1, 2, 
  4), (-2, -2, -3)\}\label{2303rays}
\ea
\ee
and 64 3d cones:
\be
\ba
\{\sigma_3\}=&\{(12, 26, 25), (12, 1, 26), (13, 23, 12), (12, 23, 9), (11, 3, 
  15), (8, 3, 18), \cr
&(19, 32, 10), (7, 31, 2), (2, 31, 6), (11, 29, 
  16), (11, 23, 13), (8, 29, 15),\cr
& (6, 29, 8), (4, 31, 9), (4, 24, 
  20), (20, 24, 11), (18, 21, 8), (21, 2, 8), \cr
&(5, 26, 19), (11, 24, 
  23), (23, 24, 9), (9, 24, 4), (34, 20, 16), (34, 28, 20),\cr
& (4, 28, 
  31), (31, 28, 6), (7, 17, 10), (10, 17, 5), (25, 26, 5), (15, 29, 
  11), \cr
&(14, 27, 10), (10, 32, 14), (28, 22, 29), (29, 22, 16), (26, 1,
   27), (30, 28, 34),\cr
& (30, 22, 28), (34, 22, 30), (27, 32, 26), (26, 
  32, 19), (16, 20, 11), (15, 3, 8), \cr
&(8, 2, 6), (13, 3, 11), (16, 22, 
  34), (27, 33, 32), (32, 33, 14), (14, 33, 27),\cr
& (27, 31, 7), (27, 1, 
  31), (10, 27, 7), (5, 19, 10), (5, 21, 18), (20, 28, 4),\cr
& (6, 28, 
  29), (7, 2, 17), (17, 21, 5), (17, 2, 21), (31, 1, 9), (25, 3, 
  12), \cr
&(9, 1, 12), (12, 3, 13), (5, 3, 25), (18, 3, 5)\}\,.
\ea
\ee

After the full blow-up to $h^{1,1}(B_{\rm toric})$, the non-Higgsable gauge group is
\be
G_{\rm nH}=E_8^{34}\times F_4^{96}\times G_2^{256}\times SU(2)^{384}\,.
\ee
There are 22 $E_8$ divisors with a non-toric $(4,6)$-curve, which are all irreducible. From $B_{\rm toric}$ to $B_3$, one needs to blow up these non-toric curves. Hence the total $h^{1,1}(X_4)$ is
\be
\ba
h^{1,1}(X_4)&=h^{1,1}(B_{\rm toric})+\mathrm{rank}(G_{\rm nH})+22+1\cr
&=3878\,.\label{3878h}
\ea
\ee

In fact, the dual polytope of (\ref{2303rays}) is exactly the convex hull of rays on $\mb{P}^3$. The Hodge number (\ref{3878h}) exactly equals to the $h^{3,1}$ of the generic elliptic CY4 over $\mb{P}^3$.

}

\item{Toric seed threefold $B_{\rm seed}$ for $h^{1,1}(B_{\rm toric})=2591$

There are 38 1d rays in $B_{\rm seed}$:
\be
\ba
\{v_i\}=&\{(0, 0, 1), (0, 1, 0), (1, 0, 0), (1, 1, 1), (1, 1, 2), (1, 2, 2), (1,
   3, 2), \cr
&(1, 2, 1), (-1, 0, -1), (0, 0, -1), (1, 1, 0), (0, -1, 
  0), (1, 0, 1), (0, -1, 1),\cr
& (1, -1, 0), (1, -1, 2), (1, -2, 
  0), (0, -1, -2), (0, 1, 1), (1, 1, 3), (1, -1, -2),\cr
& (1, -2, 
  3), (1, -2, 2), (1, 3, 3), (1, 0, -1), (1, 0, 
  2), (1, -2, -1), (1, -1, 1), \cr
&(1, -2, 1), (1, 2, 3), (1, -2, -3), (1,
   4, 3), (0, -1, -1), (1, -1, -1), (1, 0, 3),\cr
& (1, -1, 
  3), (1, -2, -2), (0, 2, 1)\}\label{2591rays}
\ea
\ee
and 72 3d cones:
\be
\ba
\{\sigma_3\}=&\{(15, 29, 28), (3, 34, 15), (2, 11, 8), (8, 11, 4), (9, 10, 2), (16, 
  36, 35),\cr
& (35, 36, 1), (4, 11, 3), (26, 35, 5), (25, 34, 3), (19, 20,
   1), (18, 33, 27), \cr
&(9, 18, 10), (9, 33, 18), (24, 32, 7), (24, 38, 
  32), (32, 38, 7), (3, 13, 4),\cr
& (1, 14, 9), (9, 14, 12), (9, 19, 
  1), (28, 29, 16), (14, 29, 12), (21, 25, 10), \cr
&(21, 34, 25), (16, 29,
   23), (23, 29, 14), (16, 35, 26), (12, 33, 9), (2, 25, 11),\cr
& (11, 25,
   3), (12, 29, 17), (17, 29, 15), (1, 36, 14), (16, 23, 22), (22, 23,
   14), \cr
&(10, 25, 2), (14, 36, 22), (22, 36, 16), (16, 26, 13), (19, 
  30, 20), (20, 30, 5),\cr
& (13, 28, 16), (9, 38, 19), (2, 38, 9), (27, 
  34, 21), (4, 26, 5), (13, 26, 4),\cr
& (15, 27, 17), (17, 27, 12), (27, 
  33, 12), (15, 34, 27), (3, 28, 13), (15, 28, 3),\cr
& (5, 35, 20), (20, 
  35, 1), (7, 8, 6), (7, 38, 8), (21, 37, 27), (27, 37, 18),\cr
& (6, 24, 
  7), (8, 38, 2), (18, 31, 21), (18, 37, 31), (31, 37, 21), (18, 21, 
  10), \cr
&(6, 8, 4), (6, 30, 24), (30, 38, 24), (5, 6, 4), (19, 38, 
  30), (5, 30, 6)\}\,.
\ea
\ee

After the full blow-up to $h^{1,1}(B_{\rm toric})$, the non-Higgsable gauge group is
\be
G_{\rm nH}=E_8^{38}\times F_4^{108}\times G_2^{288}\times SU(2)^{432}\,.
\ee
There are 22 $E_8$ divisors with a non-toric $(4,6)$-curve, which are all irreducible. From $B_{\rm toric}$ to $B_3$, one needs to blow up these non-toric curves. Hence the total $h^{1,1}(X_4)$ is
\be
\ba
h^{1,1}(X_4)&=h^{1,1}(B_{\rm toric})+\mathrm{rank}(G_{\rm nH})+22+1\cr
&=4358\,.\label{4358h11}
\ea
\ee

The low bound of the number of flps is given by (\ref{flop-estimation}):
\be
N(\mathrm{flp})\gtrsim 10^{681}\,.\label{2591flops}
\ee

Note that the polytope (\ref{2591rays}) has vertices
\be
\{V\}=\{(1,4,3),(1,-2,3),(1,-2,-3),(-1,0,-1)\}\,,
\ee
which is exactly isomorphic to the maximal reflexive polytope $S_1$ in \cite{Halverson:2017ffz,Halverson:2017vde}. 

The dual polytope of (\ref{2591rays}) has vertices
\be
\{V^\circ\}=\{(-1,0,0),(2,0,-1),(0,-1,1),(1,1,0)\}\,.
\ee
Along with a ray $(1,0,0)$, they form the rays of the generalized Hirzebruch threefold $\tilde{\mb{F}}_3$. The generic elliptic fibration over $\tilde{\mb{F}}_3$ exactly has the Hodge numbers $(h^{1,1},h^{3,1})=(3,4358)$, which is the mirror of (\ref{4358h11}).
}
\end{enumerate}

\bibliographystyle{JHEP}
\bibliography{F-ref}

\end{document}